\documentclass[12pt]{article}
\usepackage{filecontents}
\usepackage{amsmath,amsthm,amssymb}
\usepackage{mathtools}
\usepackage{graphicx,psfrag,epsf,booktabs,array}
\usepackage[title]{appendix}
\usepackage{pgfplots}
\pgfplotsset{compat=1.11}
\usepackage{subcaption}
\usepackage{enumerate}
\usepackage{etoolbox}
\usepackage{url} 
\usepackage{algpseudocode}
\usepackage{algorithm}
\usepackage{float}
\usepackage{enumitem}
\usepackage{bbm}
\usepackage{color}
\usepackage{changes}
\usepackage{natbib}
\usepackage[colorlinks=true,citecolor=blue,linkcolor=.]{hyperref}
\usepackage{soul}
\newtheorem{definition}{Definition}
\newtheorem{theorem}{Theorem}

\newtheorem{lemma}{Lemma}
\newtheorem{lmm}{Lemma A\ignorespaces}

\newtheorem{assumption}{Assumption}
\newenvironment{assumptionp}[1]{
  \renewcommand\theassumption{#1}
  \assumption
}{\endassumption}
\newcommand{\floor}[1]{\left\lfloor #1 \right\rfloor}

\usepackage{xr}
\makeatletter
\newcommand*{\addFileDependency}[1]{
\typeout{(#1)}
\@addtofilelist{#1}
\IfFileExists{#1}{}{\typeout{No file #1.}}
}\makeatother

\newcommand*{\myexternaldocument}[1]{%
\externaldocument{#1}%
\addFileDependency{#1.tex}%
\addFileDependency{#1.aux}%
}
\myexternaldocument{supplementary}

\newcommand{\E}{\mathrm{E}}
\newcommand{\Var}{\mathrm{Var}}

\algnewcommand{\Inputs}[1]{%
  \State \textbf{Inputs:}
  \Statex \hspace*{\algorithmicindent}\parbox[t]{.8\linewidth}{\raggedright #1}
}
\algnewcommand{\Initialize}[1]{%
  \State \textbf{Initialize:}
  \Statex \hspace*{\algorithmicindent}\parbox[t]{.8\linewidth}{\raggedright #1}
}

\newcommand{\blind}{1}

\date{}

\begin{document}

\def\spacingset#1{\renewcommand{\baselinestretch}%
{#1}\small\normalsize} \spacingset{1}


\if1\blind
{
   \title{\bf
  Independence-Encouraging Subsampling for Nonparametric Additive Models
  }
  \author{Yi Zhang\\
    Department of Statistics, George Washington University\\
    Lin Wang \\
    Department of Statistics, Purdue University\\
    Xiaoke Zhang 
    \\
    Department of Statistics, George Washington University\\
    and\\
    HaiYing Wang\\
    Department of Statistics, University of Connecticut
    }
  \maketitle
} \fi

\if0\blind
{
  \bigskip
  \bigskip
  \bigskip
  \begin{center}
    {\LARGE\bf Independence-Encouraging Subsampling for Nonparametric Additive Models}
\end{center}
  \medskip
} \fi

\bigskip
\begin{abstract} 
The additive model is a popular nonparametric regression method due to its ability to retain modeling flexibility while avoiding the curse of dimensionality. The backfitting algorithm is an intuitive and widely used numerical approach for fitting additive models. However, its application to large datasets may incur a high computational cost and is thus infeasible in practice. To address this problem, we propose a novel approach called independence-encouraging subsampling (IES) to select a subsample from big data for training additive models. Inspired by the minimax optimality of an orthogonal array (OA) due to its pairwise independent predictors and uniform coverage for the range of each predictor, the IES approach selects a subsample that approximates an OA to achieve the minimax optimality. Our asymptotic analyses demonstrate that an IES subsample converges to an OA and that the backfitting algorithm over the subsample converges to a unique solution even if the predictors are highly dependent in the original big data. The proposed IES method is also shown to be numerically appealing via simulations and a real data application. 
\end{abstract}

\noindent%
\vfill

\newpage
\spacingset{1.45} 
\section{Introduction}
\label{sec:intro}
Big data of huge sample sizes are prevalent in many disciplines such as science, engineering, and medicine. Such data may reveal important domain knowledge, but meanwhile they pose challenges to data storage and analysis. To address those challenges, subsampling has recently received increasing attention and has been intensively studied.

An optimal subsampling approach typically specifies a downstream model and
carefully selects an informative subsample so that the model training on the
subsample is more accurate than that on other possible subsamples. 
Different subsampling approaches have been developed for various parametric models.
For linear regression, \citet{ma} proposed subsampling probabilities defined via leverage scores. \citet{wang_information-based_2019} investigated an information based optimal subsampling algorithm motivated by $D$-optimal experimental design.
\citet{lin_wang_oabdlr_2021} developed an orthogonal subsampling (OSS) method inspired by the universal optimality of orthogonal array (OA) for linear regression. 
Subsampling methods for other parametric models are also extensively studied, such as \citet{wang_optimal_2018} and \cite{han2020local} for logistic regressions,
\citet{wang2021optimal} for quantile regression, and \citet{ai_optimal_2021} for
generalized linear models. Despite their optimality in some sense for fitting specific parametric models, the usage of those methods can be
hindered by strong model assumptions that may not hold in big data problems. See
\citet{fan2014challenges} for a detailed discussion. To this end, 
\citet{meng_lowcon_2021} proposed an algorithm, called LowCon, to select a
space-filling subsample which is shown to be robust when a linear model is misspecified. 
Researchers have also looked into nonparametric settings with less stringent
model assumptions. For example, \citet{ma_eff_spline_spacefilling_2020} showed the superiority of a space-filling subsample for multivariate smoothing splines; \citet{kernel_sketch_2017} applied tensor sketching to accelerate kernel ridge regression; \citet{zhao2018efficient} and
\citet{he2022gaussian} considered design-based subsampling for
Gaussian process modeling; \citet{shi2021model} considered model-robust subdata selection.
Other methods include
continuous distribution compression \citep{Simon_supportpoints_2018} and supervised data
compression \citep{joseph2021supervised}.

The nonparametric additive model \citep{hastie_tibshirani_additive_1986} has been widely used in practice because of its interpretability and flexibility \citep[e.g.,][]{walker2002comparing,hwang2009additive,liutkus2014kernel}. It avoids the ``curse of dimensionality" which impedes the implementation of fully nonparametric models with multiple predictors. However, fitting an additive model may still be computationally expensive when the sample size is huge. 
For example, if the backfitting algorithm \citep{breiman1985estimating,buja1989linear} combined with local polynomial smoothing is used 
to fit an additive model on a data set with $N$ observations of $p$ predictors,
where $p\ll N$, 
the time complexity is $O(N^2)$ per backfitting iteration. If the bandwidth is
selected via cross-validation, then the complexity would become $O(N^2)$ per
bandwidth grid evaluation. Therefore, the practicality of additive models is hindered for large data.

We propose an independence-encouraging subsampling (IES) method for fitting an additive model with big data. Akin to the OSS \citep{lin_wang_oabdlr_2021}, the IES is inspired by the robustness and optimality of OA for experimental design and data collection \citep{cheng1980orthogonal, taguchi1990robust}. Nevertheless, existing results for OAs focus on their optimality for identifying main effects and interactions via linear regression. 
We first
derive theoretical results on the minimax optimality of random
OAs for nonparametric additive models and then develop the IES method to select a
subsample that approximates a random OA. The merits of IES are three-fold. Firstly, it is fast and easy to implement. The computation of selecting a subsample and training a nonparametric additive model on the subsample is significantly faster than training the model on the large full
data. Secondly, our theoretical analyses show that an IES subsample converges to
a random OA whose predictors achieve marginal uniformity and pairwise independence.
This substantially benefits the backfitting algorithm, the most popular numerical approach to fit additive models. A well-known sufficient condition for local polynomial backfitting estimator to converge is the ``near independence" between predictors \citep{opsomer_fitting_1997}.
Since the predictors are empirically independent in the selected subsample, the
nbackfitting algorithm will converge to a unique solution even if the
predictors are highly dependent in the original big data.
Lastly, the IES approach is numerically shown to be superior to existing subsampling methods for fitting additive models and robust against certain model misspecifications.

The remainder of this paper proceeds as follows. 
Section \ref{sec:sam} derives the minimax optimal sampling plan for additive
models. Section \ref{sec:roas} introduces random OAs and their properties. Section \ref{sec:IES} proposes the IES subsampling approach, and develops some asymptotic theories. Section \ref{sec:PI} provides a fast implementation algorithm for IES. 
Sections \ref{sec:simu} and \ref{sec:real} present simulations and a real data
example, respectively. Discussion in Section \ref{sec:conc} concludes this paper. 
Technical proofs are provided in the Supplementary Materials. R code is publicly available at \url{https://github.com/....}. 



\section{Minimax Optimal Sampling Plan}
\label{sec:sam}




In this section, we introduce the minimax optimal sampling plan for univariate nonparametric regression and then extend it to additive models. 

\subsection{
Optimal sampling for
univariate nonparametric regression}\label{sec:uni}

We first consider univariate nonparametric regression for independent and identically
distributed (i.i.d.) data: 
\begin{equation}\label{eq:univar}
Y_i=m(X_i)+\epsilon_i, \quad i=1,\ldots, N,
\end{equation}
where for the $i$-th subject, $i=1,\ldots, N$, $X_i$ is the univariate
continuous predictor, $Y_i$ is the response, $\epsilon_i$ is the random error,
and $m(x) = E(Y_i \mid X_i =x)$ 
is the regression function.
The support of the predictor is assumed 
compact and hereafter $[0,1]$ without loss of generality.
It is also assumed that $\epsilon_i$ are
independent of the predictors, 
$\E(\epsilon_i)=0$, and $\Var(\epsilon_i)=\sigma^2$. 

The literature of univariate nonparametric regression \citep[e.g., Chapter 5 of][]{wasserman2006all} favors linear smoothers of the form
\begin{equation}\label{linearS}
\tilde{m}(x)=\sum_{i=1}^{N}w_i(x;X_1,\dots,X_N)Y_i,
\end{equation}
where $w_i$ is a data-dependent weight function.
Among them, the local linear estimator is a popular option. \citet{fan_desada_1992} showed that under mild conditions, the local linear estimator for 
\eqref{eq:univar} asymptotically achieves a minimax risk on the mean squared
error (MSE), where the minimum is taken over all linear smoothers and the maximum is taken over all $m(\cdot)$ in 
\begin{equation}\label{eq:1}
\mathcal{C}^*=\left\{m(x)\in
  C^{(2)}[0,1]\quad\big|\quad\max_x|m^{(2)}(x)|^2\le\eta\right\},
\end{equation}
with $C^2[0,1]$ denoting the set of
functions whose second derivatives are continuous. For any $x\in[0,1]$, the minimax risk is 
$$
R_0(x)=\frac{3}{4}15^{-1/5}\left\{ \frac{\eta^{1/4}\sigma^2}{Nf(x)} \right\}^{4/5} \{1+o_P(1)\},
$$
where $f$ is the density of the predictor distribution. 

The $R_0(x)$ may still be large for the region with a small
$f(x)$. We hope that an estimator is ``robust" for all $m(\cdot)$ in $\mathcal{C}^*$ and all $x\in[0,1]$,
in the sense that the estimator performs well even
in the worst scenario. 
Therefore, we seek a sampling regime, or equivalently a design density $f$,
that minimizes the following minimax risk:  
\begin{align}\label{eq:modmm}
R(f)=\min_{\tilde{m}(x) \mbox{\footnotesize ~linear}}\sup_{m\in \mathcal{C}^*,x\in[0,1]}\E[(\tilde{m}(x)-m(x))^2 \mid X_1, \dots, X_N],
\end{align}
where $\min_{\tilde{m}(x) \mbox{\footnotesize ~linear}}$ takes the minimum over all linear smoothers in \eqref{linearS}, and $\mathcal{C}^*$ is defined in \eqref{eq:1}.
The following result calculates the $R(f)$ in \eqref{eq:modmm} and
provides the optimal $f$ that minimizes $R(f)$. Denote $[a]_+=\max\{0,a\}.$

\begin{theorem}\label{l1}
Suppose that $f(x)$ is bounded away from zero and infinity. Let $f(x_0)=\min_{x\in[0,1]}f(x)$. The minimax risk in \eqref{eq:modmm} is given by
\begin{equation}\label{eq:eq1}
 R(f)=\frac{3}{4}15^{-1/5}\eta^{1/5}\left(\frac{\sigma^2}{Nf(x_0)}\right)^{4/5} (1+o_p(1)),
\end{equation}
which is achieved by the local linear regression estimator with the Epanechnikov kernel $ K_0(u)=3[1-u^2]_{+}/4$ and bandwidth $h_0=\left\{15\sigma^2/[N\eta f(x_0)]\right\}^{1/5}.$ The optimal design density $f$ that minimizes $R(f)$ in \eqref{eq:eq1} is the uniform density, that is, $f(x)=1$ for all $x\in[0,1]$.
\end{theorem}



 
\subsection{
Optimal sampling for
additive models}\label{sec:add}
We now consider an additive model for i.i.d. data:
\begin{equation} \label{eq:addm}  Y_i=
 m(\mathbf{X}_i) +\epsilon_i=\mu+m_1(X_{i1})+m_2(X_{i2})+\dots+m_p(X_{ip})+\epsilon_i, \quad i=1,2,\dots,N,
 \end{equation}
where 
$\mathbf{X}_i=(X_{i1},\dots,X_{ip})$ contains $p$ predictors,  
$Y_i$ is the response, $m(x) = E(Y_i | \mathbf{X}_i = x)$ is the regression function,  
$\mu$ is a constant, 
$m_j(x)$ is the component function for the $j$-th predictor assumed to be smooth, 
and $\epsilon_i$'s are random errors. 
Again, the support of each predictor is assumed $[0,1]$ without loss of
generality, and $\epsilon_i$ is independent of the predictors with 
$\E(\epsilon_i)=0$ and $\Var(\epsilon_i)=\sigma^2$. 
Moreover, the following condition is imposed for identifiability: 
\begin{equation}\label{eq:indencon}
\int_{0}^{1}m_j(x)\,dx=0, \quad j=1, \ldots, p.
\end{equation}



The backfitting algorithm
\cite[e.g.,][]{breiman1985estimating,buja1989linear} is a popular, intuitive, and easy-to-implement numerical approach for fitting additive models. 
The algorithm updates each component function estimator alternately and iteratively. 
At each iteration, a one-dimensional smoother, e.g., the local linear smoother, is
applied to regress the residual on one predictor to update its corresponding
component function estimate, where the residual is obtained by subtracting all
the other component functions' estimates from the response.
The asymptotic properties of the backfitting algorithm 
have
been studied 
by \cite{opsomer_fitting_1997} and \cite{opsomer_asymptotic_2000}. 
Their results also indicate that the convergence of the backfitting algorithm is not theoretically guaranteed if some predictors are highly dependent.



On the contrary, if all predictors are pairwise independent, 
\eqref{eq:addm} implies that $$m_{j}(X_{ij})-\E [m_{j}(X_{ij})]=\E[Y_i \mid X_{ij}] - \E[Y_i], \quad \text{for each $j=1,\ldots, p$,}
$$ 
where the left-hand side is a centered component function and the right-hand side suggests a univariate
regression of the response on the $j$-th predictor. 
Hence, pairwise independence separates the additive modeling problem to $p$ one-dimensional estimations, so no iteration is required. 
{In fact, as shown in \citet{opsomer_fitting_1997}, ``near independence" between predictors can ensure the local-polynomial-based backfitting algorithm to converge.
Therefore, inspired by Theorem \ref{l1}, we recommend sampling predictors independently and uniformly to achieve the minimax optimality for each component function estimation. 
By Theorem 3.1 of \citet{opsomer_asymptotic_2000}, marginal uniformity is also optimal in minimizing the conditional variance of each local polynomial-based backfitted component function estimator over all possible designs.}


When selecting a subsample from large data, since the data may have highly dependent predictors and follow an arbitrary distribution, obtaining a subsample with independently and uniformly distributed predictors (at the population level) is typically impossible. However, we can seek empirical independence and uniformity for predictors in the subsample, and this can be achieved via random OA.

\section{Introduction to OA}\label{sec:roas}

An OA of strength $t$, denoted by OA$(N,p,q,t)$, is an $N\times p$ matrix with entries of $q$ levels indexed by $\{0,1,2,\dots,q-1\}$, arranged in such a way that all level combinations occur equally often in any $t$ columns  \citep{hedayat1999orthogonal}. Such equal frequency of level combinations is called combinatorial orthogonality. The following matrix, as an example, is an OA$(4,3,2,2)$, any two columns of which consist of $(0,0)$, $(0,1)$, $(1,0)$, and $(1,1)$ exactly once: 
\begin{equation}\label{oaeg1}
\begin{pmatrix}
0 & 0 & 0 \\
0 & 1 & 1 \\
1 & 0 & 1 \\
1 & 1 & 0 \\
\end{pmatrix}. 
\end{equation}
In this paper, OAs mentioned are assumed to have strength 2 unless otherwise specified. 

OAs have been extensively used as fractional factorial designs because they allow uncorrelated estimation of main effects through linear regression \citep{wu2011experiments, mukerjee2006modern,wang2022class}.  \citet{cheng1980orthogonal} showed that an OA on $q$ levels is universally optimal, i.e., optimal under a wide variety of criteria that include $D$- and $A$-optimality, among all $q$-level factorial designs for studying main effects. 

We now extend the superiority of OAs for establishing nonparametric additive models. 
Consider the sampling distribution of the column variables $A_j$ in an OA. We have $P(A_j=a)=q^{-1}$ and $P(A_j=a, A_{j'}=a')=P(A_j=a)P(A_{j'}=a')=q^{-2}$ for all $a, a'\in\{0,1,2,\dots,q-1\}.$ Therefore, any column variable in an OA follows a discrete uniform distribution, and any pair of column variables are independent. 
We next provide a sampling scheme to draw data from $[0,1]^p$ that carry over the uniformity and variable independence of an OA.
\begin{definition}
Given an OA$(N,p,q,2)$, denoted by $\mathcal{A}=(a_{ij})$ for $i=1,\ldots,N$ and $j=1,\ldots,p$, a random OA $(X_{ij})$ is given by
$$X_{ij}=\frac{a_{ij}+U_{ij}}{q}, \mbox{ for } i=1,\ldots,N, \mbox{ and } j=1,\ldots,p,$$
where the $U_{ij}$'s are independent uniform random variables on $[0,1]$. 
\end{definition}

\begin{figure}[t]
\centering
\includegraphics[scale=0.5]{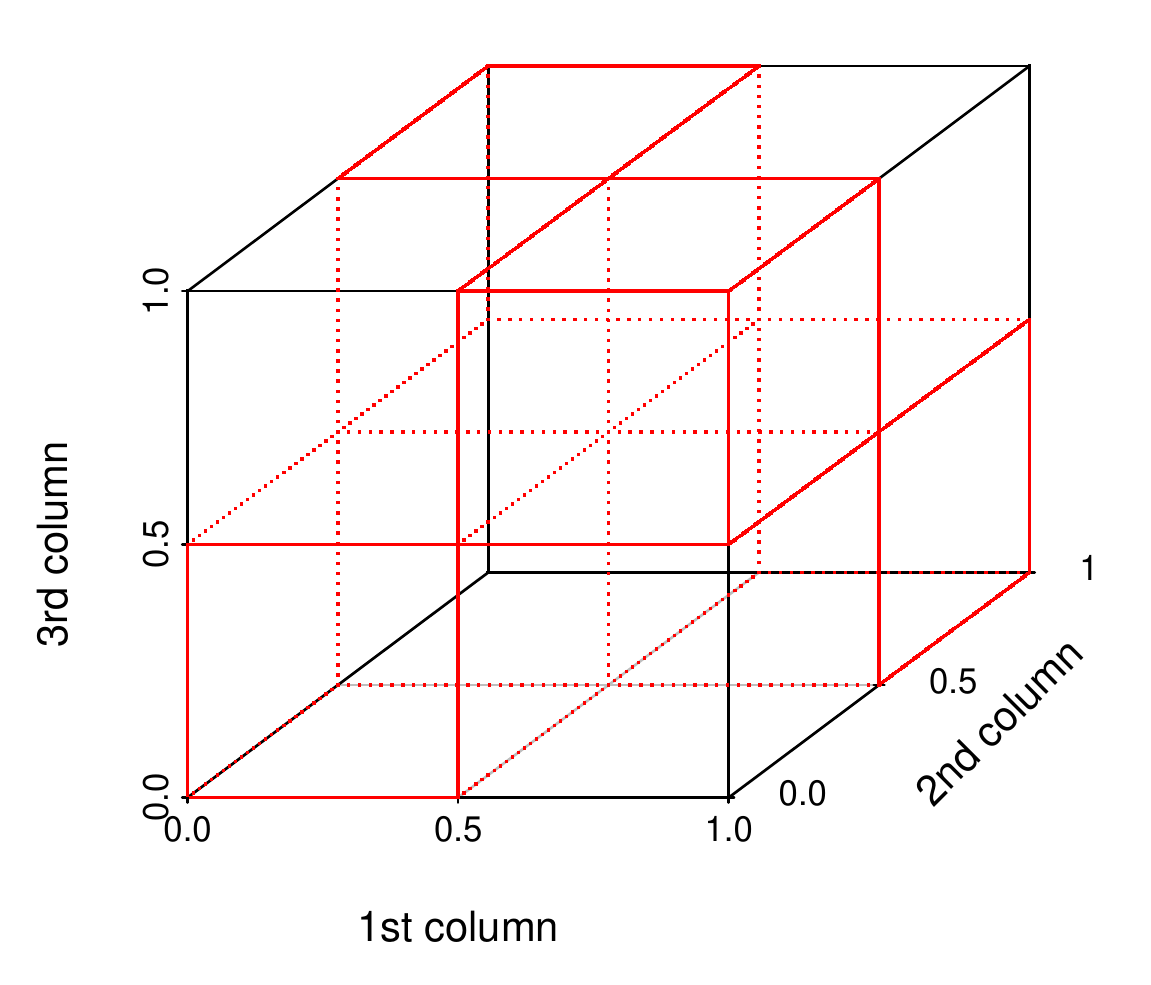}
\caption{Illustration of selected cells given by \eqref{oaeg1}. A cell is selected if its all edges are red.} 
\label{fig:eg2}
\end{figure}

A random OA can be understood as a two-step sampling procedure. Firstly, partition the cube $\lbrack 0, 1\rbrack^p$ into $q^p$ equal-sized cells (subcubes with each side of length $q^{-1}$) and select the $n$ cells specified by the rows of $\mathcal{A}$. The $i$th row of $\mathcal{A}$ specifies the cell $\Pi_{j=1}^{p}[a_{ij}/q,(a_{ij}+1)/q)$. Secondly, randomly draw a point from each selected cell. 
Figure \ref{fig:eg2} illustrates the four selected cells according to \eqref{oaeg1}. For any two columns, the projection of selected cells covers the whole face. Therefore, the randomly sampled points from those cells uniformly cover any two-dimensional subspace. Such a sampling scheme was also studied in \citet{owen1992orthogonal} to obtain a better approximation of integration than Monte Carlo sampling. 


\begin{lemma}\label{UOASdist}
For a random OA, the cumulative distribution on each column is given by $F(x_1)=x_1$, and on any pair of columns is given by $F(x_1,x_2)=x_1x_2$.
\end{lemma}
Lemma~\ref{UOASdist} claims both uniformity and pairwise independence between column variables in a random OA, which are inherited from its combinatorial orthogonality and are the exact properties we seek for the optimal training data for additive models. 
It should be noted that for an OA$(N,p,q,2)$ to exist, the number of rows has to be a multiple of $q^2$, that is, $N=\lambda q^2$ for some positive integer $\lambda$. Abundant methods have been proposed to generate OAs, and we relegate a summary of their wide availability and generating methods to Appendix \ref{appenda}.

\section{Independence-Encouraging Subsampling (IES)}\label{sec:IES}
Let $(\mathbf{x}_1,y_1),\dots,(\mathbf{x}_N,y_N)$ denote the full data with $N$ observations, where
$\mathbf{x}_i=(x_{i1},\ldots,x_{ip})$ are observations of $p$ predictors and $y_i$ is the corresponding response. We consider taking a subsample of size $n$, denoted as $(\mathbf{x}_1^*,y_1^*),\dots,(\mathbf{x}_n^*,y_n^*)$.
Based on the previous discussion, our goal is to encourage empirical uniformity and pairwise independence of predictors in the subsample, and this can be achieved by finding a subsample whose design matrix approximates a random OA. 

An intuitive approach is to choose an existing OA with $n$ rows and randomly select a data point in each cell specified by the OA. 
This approach has two possible limitations. First, for an OA$(n,p,q,2)$ to exist,
the number of rows has to be a multiple of $q^2$, meaning that this approach is possible only when $n=\lambda q^2$ for some positive integer $\lambda$. 
Second, even if $n$ is a multiple of $q^2$, the full data
may not fit an arbitrarily chosen OA, that is, many cells of the OA may be empty and do not contain any data points. 

The proposed IES method selects a subsample by directly minimizing a discrepancy
function that measures its deviation from an OA. As a result, the subsample size is not restricted to be a multiple of a square number, and the selected subsample approximates an OA that is the best
compatible with the data.


\subsection{The IES approach}
For a full data with design matrix $\mathcal{X}=(x_{ij})$ and a prespecified integer $q$,
define the membership matrix as $\mathcal{Z}=(z_{ij})$,
where $$z_{ij}=\floor{x_{ij}q},$$ for $i=1,2,\dots,N$, and $j=1,2,\dots,p.$
Clearly $z_{ij}\in\{0,1,2,\dots,q-1\}$.
Our goal is to search for a subsample whose design matrix $\mathcal{X}^*$ has an
OA membership matrix.
For any two observations with $\mathbf{x}_i$ and $\mathbf{x}_{i'}$, define
\[ \delta(\mathbf{x}_i,\mathbf{x}_{i'})=\sum_{j=1}^{p}\mathbbm{1}(\lfloor x_{ij}q\rfloor=
\lfloor x_{i'j}q\rfloor)= \sum_{j=1}^{p}\mathbbm{1}(z_{ij}=z_{i'j}),\]
where $\mathbbm{1}{(z_{ij}=z_{i'j})}$ is the indicator function that equals 1 if
$z_{ij}=z_{i'j}$ and 0 otherwise. Here, $\delta(\mathbf{x}_i,\mathbf{x}_{i'})$
counts the membership coincidence between elements of $\mathbf{z}_i$ and $\mathbf{z}_{i'}$, and thus
measures the similarity between $\mathbf{x}_i$ and $\mathbf{x}_{i'}$.  For a
subsample with design matrix
$\mathcal{X}^*=(\mathbf{x}^*_1,\dots,\mathbf{x}^*_n)^{T}$, define 
\begin{equation}\label{Ldef}
    L(\mathcal{X}^*)=\sum_{1\leq i< i'\leq n}\lbrack\delta(\mathbf{x}_i^*,\mathbf{x}_{i'}^*)\rbrack^2.
\end{equation}
Clearly, $L(\mathcal{X}^*)$ measures the overall similarity between all data points in $\mathcal{X}^*$. 
The following theorem shows that $L(\mathcal{X}^*)$ also measures the discrepancy between $\mathcal{X}^*$ and an OA. 

\begin{theorem}\label{oalemma}
For any $\mathcal{X}^*$ with
$n$ rows, $$L(\mathcal{X}^*)\ge\frac{n}{2q^2}[np(p+q-1)-(pq)^2],$$ and the lower
bound is achieved if and only if $\mathcal{Z}$, the membership matrix of $\mathcal{X}^*$, is an OA$(n,p,q,2)$.
\end{theorem}

Theorem \ref{oalemma} shows that $L(\mathcal{X}^*)$ has a lower bound which is attained if and
only if the membership matrix of $\mathcal{X}^*$ forms an OA. In this sense, $L(\mathcal{X}^*)$ can be viewed as a metric on the discrepancy between $\mathcal{X}^*$ and a realization of a random OA. 
Therefore, we propose the IES method, which solves the
optimization problem:
\begin{equation}\label{minL}
    \mathcal{X}^{*}_{opt}=\arg \min_{\mathcal{X}^{*}\subset\mathcal{X}} L(\mathcal{X}^{*}).
\end{equation}
The IES subsample is $\{\mathcal{X}^{*}_{opt},\mathbf{y}^{*}_{opt}\}$, where $\mathbf{y}^{*}_{opt}$ is the corresponding response vector.

The optimization in \eqref{minL} 
does not impose any restriction on $n$. When $n=\lambda q^2$ and an OA$(n,p,q,2)$ exists,
we obtain a subsample from \eqref{minL} with an OA
membership matrix. Otherwise, we obtain a subsample that approximates the
combinatorial orthogonality in an OA. 
We can extend Theorem \ref{oalemma} to a more general setting of $n$ for which an OA$(n,p,q,2)$
may not exist, which confirms that the optimization in \eqref{minL} best approximates an OA for a general setting of $n$. The presentation of the result requires tedious notations and concepts, so we relegate the details to Lemma S\ref{subsampwoa} in Supplementary Material. 

We next investigate the asymptotic properties of an IES subsample selected by
\eqref{minL}, under the following assumptions.

\begin{assumptionp}{1}\label{a2}
The probability density function that generates the design matrix of the full data is compactly supported and bounded away from
zero and infinity.
\end{assumptionp}


\begin{assumptionp}{2}\label{a1}
  There exists some fixed positive integer $\lambda$ such that
$n-\lambda q^2=O(q)$, and an OA$(q^2, p+1, q, 2)$ exists. 
\end{assumptionp}

\begin{assumptionp}{3}\label{a3}
The subsample size $n$ goes to $\infty$ at the rate of $O(N^{\nu})$ for some $\nu
\in\left(0,2/p\right)$.
\end{assumptionp}

Assumption~\ref{a2} ensures that the full data asymptotically cover the design region as the size $N$ increases.
Assumption~\ref{a1} indicates again that the IES does not require $n=\lambda q^2$. 
The requirement of the existence of OA$(q^2, p+1, q,2)$ is weak, as discussed in Appendix \ref{appenda}, especially considering that we can set $q$ to be much bigger than $p$.  
Assumption \ref{a3} requires that $n$ does not grow faster than $N^{2/p}$, which
is commonly the case in the setting of big subsampling.

\begin{theorem}\label{thm:inp} 
Define the induced joint cumulative distribution function on any two columns of $\mathcal{X}^{*}_{opt}$, $X_{j}^*$ and $X_{j'}^*$, as $$F_{n}(x_1,x_2)=\frac{1}{n}\sum_{i=1}^{n}\mathbbm{1}(X^*_{ij}\le x_1, X^*_{ij'}\le x_2).$$ 
Then under Assumptions \ref{a2}-\ref{a3}, we have $$\sup_{x_1,x_2\in[0,1]}\left|F_{n}(x_1,x_2)-x_1x_2\right|=O_p\left(N^{-\nu/2}
\right).$$
\end{theorem}


Theorem \ref{thm:inp} shows that asymptotically the solution to \eqref{minL} achieves pairwise independence and uniformity, leading to a desired subsample for additive models. The convergence rate depends on $\nu$ in Assumption \ref{a3}. A bigger $\nu$ indicates a larger subsample size and results in a faster convergence to the uniform distribution. We can relax Assumption \ref{a1} to a more general setting of $n$ with $n-\lambda q^2=O(q^{\gamma})$ for some $\gamma\in(0,2)$. The case of $\gamma\leq1$ is equivalent to Assumption \ref{a1}, and for $\gamma>1$, $F_{n}(x_1,x_2)$ still converges to uniformity but at a slower rate; 
see the proof of Theorem \ref{thm:inp} in the Supplementary Materials for details.  

\subsection{Additive Modeling on IES Subsamples}\label{subest}

After obtaining the subsample $\{\mathcal{X}^{*}_{opt},\mathbf{y}^*_{opt}\}$ from \eqref{minL}, we fit an additive model on this subsample. Since the predictors in the subsample cannot be guaranteed to be perfectly independent, we propose to estimate each component function via the backfitting algorithm \citep{breiman1985estimating}. Motivated by Theorem \ref{l1}, we apply local linear smoothers in each backfitting step. 

When there are two predictors, i.e., $p=2$, we can prove the convergence of the backfitting algorithm on the subsample $\{\mathcal{X}^{*}_{opt},\mathbf{y}^*_{opt}\}$. 
We need the following assumptions in addition to Assumptions \ref{a2}--\ref{a3}.


\begin{assumptionp}{4}\label{a4}
The kernel function $K$ is a symmetric density function compactly supported on $[-1,1]$. Moreover, $K$ is $M$-Lipschitz for some constant $M>0$, i.e., 
$|K(u) - K(v)| \leq M |u-v|$ for any $u, v \in [-1, 1]$.
\end{assumptionp}


\begin{assumptionp}{5}\label{a5}
As the size of the subsample $n\rightarrow\infty$, the bandwidth $h_j\rightarrow 0$ and $nh_j^4\rightarrow\infty$ for $j=1,2$.
\end{assumptionp}

Both Assumptions \ref{a4} and \ref{a5} will be used in the proof of Theorem \ref{thm:asym} to control certain numerical integration errors.
Assumption \ref{a4} on the kernel function is commonly adopted by kernel-smoothing-based additive modeling methods \citep[e.g.,][]{opsomer_fitting_1997,Zhang-Park-Wang2013} and can be satisfied by popular kernels, e.g., the Epanechnikov kernel.
Assumption \ref{a5} on bandwidths is mild and can be satisfied if each $h_j$ takes the optimal order $n^{-1/5}$ as in the literature of local polynomial smoothing \citep[e.g.,][]
{fan1996local}.




\begin{theorem}\label{thm:asym}
Under Assumptions~\ref{a2}-\ref{a5},
when $p=2$, the
backfitting algorithm on the subsample 
$\{\mathcal{X}^{*}_{opt},\mathbf{y}^*_{opt}\}$ converges to a unique solution with probability
approaching one as $N\rightarrow\infty$.
\end{theorem}


The expression of the unique solution involves
more tedious notations and can be unwieldy in practice. To save space, we
defer the details to Appendix \ref{appendb}. Substantially different from the result by \citet{opsomer_fitting_1997}, Theorem \ref{thm:asym} does not require a weak dependency between the two predictors in the population; even if the population dependency between the predictors is high, they are almost independent in $\mathcal{X}^{*}_{opt}$ as guaranteed by Theorem \ref{thm:inp}, 
so the backfitting procedure on the subsample can converge asymptotically. Another critical distinction between Theorem \ref{thm:asym} and \citet{opsomer_fitting_1997} is that the latter handles independent observations while observations in $\mathcal{X}^{*}_{opt}$ are dependent.

When $p \geq 3$, theoretical convergence for the backfitting procedure on an IES subsample is unknown and will be deferred for future work. Nevertheless, it always converges numerically in our simulation studies and real data application in Sections \ref{sec:simu} and \ref{sec:real}.





\section{Practical Implementation of IES}\label{sec:PI}

The optimization problem in \eqref{minL} is computationally expensive to solve. An exhausted search requires evaluating the quantity $L(\mathcal{X}^*)$ on $\binom{N}{n}$ possible subsamples, which is prohibitive for even a moderate data size. To improve the efficiency, we propose a sequential IES implementation which selects subsample points sequentially. 
We start with a randomly selected point $(\mathbf{x}_1^*,y_1^*)$. Denote the subsample design matrix with $k$ points as $\mathcal{X}^{*}_{(k)}=(\mathbf{x}^{*}_{1},\dots,\mathbf{x}^{*}_{k})^{T}$ for $k\in\{1,\ldots, n-1\}$. 
The $(k+1)$th subsample point is then selected as
\begin{align*}
    \mathbf{x}^*_{k+1}&=\arg\min_{\mathbf{x}\in \mathcal{X}/\mathcal{X}^{*}_{(k)}}L(\mathcal{X}^{*}_{(k)}\cup{\{\mathbf{x}}\})\\
    &=\arg\min_{\mathbf{x}\in \mathcal{X}/\mathcal{X}^{*}_{(k)}}\left\{\sum_{1\leq i< i'\leq k}\lbrack\delta(\mathbf{x}_i^*,\mathbf{x}_{i'}^*)\rbrack^2+\sum_{1\leq i\leq k}[\delta(\mathbf{x}_i^*,\mathbf{x})\rbrack^2\right\}\\
    &=\arg\min_{\mathbf{x}\in \mathcal{X}/\mathcal{X}^{*}_{(k)}} l(\mathbf{x}\mid\mathcal{X}^{*}_{(k)}),
\end{align*}
where 
\begin{equation}
    l(\mathbf{x}\mid\mathcal{X}^{*}_{(k)})=\sum_{1\leq i\leq k}\lbrack \delta(\mathbf{x}_i^*,\mathbf{x})\rbrack^2
\end{equation}
measures the similarity between $\mathbf{x}$ and $\mathcal{X}^{*}_{(k)}$, and the selected $ \mathbf{x}^*_{k+1}$ is the least similar point to $\mathcal{X}^{*}_{(k)}$. If there are multiple minimizers, $ \mathbf{x}^*_{k+1}$ is randomly selected among them. 
After choosing $\mathbf{x}^{*}_{k+1}$, we update $l(\cdot)$ for $\mathbf{x}\in\mathcal{X}/\mathcal{X}^{*}_{(k+1)}$ via $$l\left(\mathbf{x} \mid \mathcal{X}_{(k+1)}^{*}\right)=l\left(\mathbf{x} \mid \mathcal{X}_{(k)}^{*}\right)+\delta(\mathbf{x},\mathbf{x}^*_{k+1})^2,$$ 
so the computational complexity of selecting one point is $O(Np)$. 

\begin{algorithm}[t]
\caption{Sequential IES Method}\label{alg:oss}
\begin{algorithmic}
\Inputs{Full data $\{\mathcal{X},\mathbf{y}\}$, subsample size $n$, hyperparameter $q$}
\Initialize{ 
 \State Set $\{\mathcal{X}^{*}_{(1)},\mathbf{y}^*_{(1)}\}\gets (\mathbf{x}_{1}^*,y_{1}^*), \text{ with $(\mathbf{x}_{1}^*,y_{1}^*)$ randomly selected }$\
\State Calculate $l\left(\mathbf{x}\mid \mathcal{X}^{*}_{(1)}\right)$, for all $\mathbf{x}\in \mathcal{X}/\mathcal{X}^{*}_{(1)}$}
\For{$k= 1$ to $n-1$} 
 \State {$\mathbf{x}_{k+1}^*\gets$ randomly sample one point from $\text{arg}\min_{x\in \mathcal{X}/\mathcal{X}^{*}_{(k)}} l\left(\mathbf{x}\mid \mathcal{X}^{*}_{(k)}\right)$} 
 \State $\{\mathcal{X}^{*}_{(k+1)},\mathbf{y}_{(k+1)}^*\}\gets \{\mathcal{X}^{*}_{(k)},\mathbf{y}_{(k)}^*\}\cup\{(\mathbf{x}^*_{k+1},y^*_{k+1})\}$
 \State $l\left(\mathbf{x}\mid \mathcal{X}^{*}_{(k+1)}\right)\gets l\left(\mathbf{x}\mid \mathcal{X}^{*}_{(k)}\right)+\delta(\mathbf{x},\mathbf{x}^{*}_{k+1})^2$, for all $\mathbf{x}\in \mathcal{X}/\mathcal{X}^{*}_{(k+1)}$
\EndFor
\State Apply a backfitting algorithm to the selected subsample $\{\mathcal{X}^{*}_{(n)},\mathbf{y}_{(n)}^*\}$
\State\Return{ $\hat{\mu}$ and $\hat{m}_j$, for $j=1,2,\dots,p$, trained with the backfitting algorithm}
\end{algorithmic}
\end{algorithm}

Algorithm~\ref{alg:oss} outlines the detailed steps of the sequential IES implementation. In our numerical results in Sections \ref{sec:simu} and \ref{sec:real}, the backfitting algorithm uses the local linear smoothing and is conducted via the R package \textit{gam} \citep{hastie2015package}. The hyperparameter $q$ can be any not-to-small integer, and we find that an integer greater than 10 would be adequate. Also, setting $q$ at a prime power may provide more stable numerical performance because of the better OA approximation and combinatorial orthogonality (details in Appendix A). 
Therefore, we recommend choosing a prime power $q$ which is close to $\sqrt{n/\lambda}$ for some positive integer $\lambda$. 
In our simulation and real data studies where $n=1000$ and $5000$, we set $q=2^4=16$, which is close to $\sqrt{1000/4}=15.8$.

\begin{figure}[t]
\centering
\includegraphics[scale=0.7]{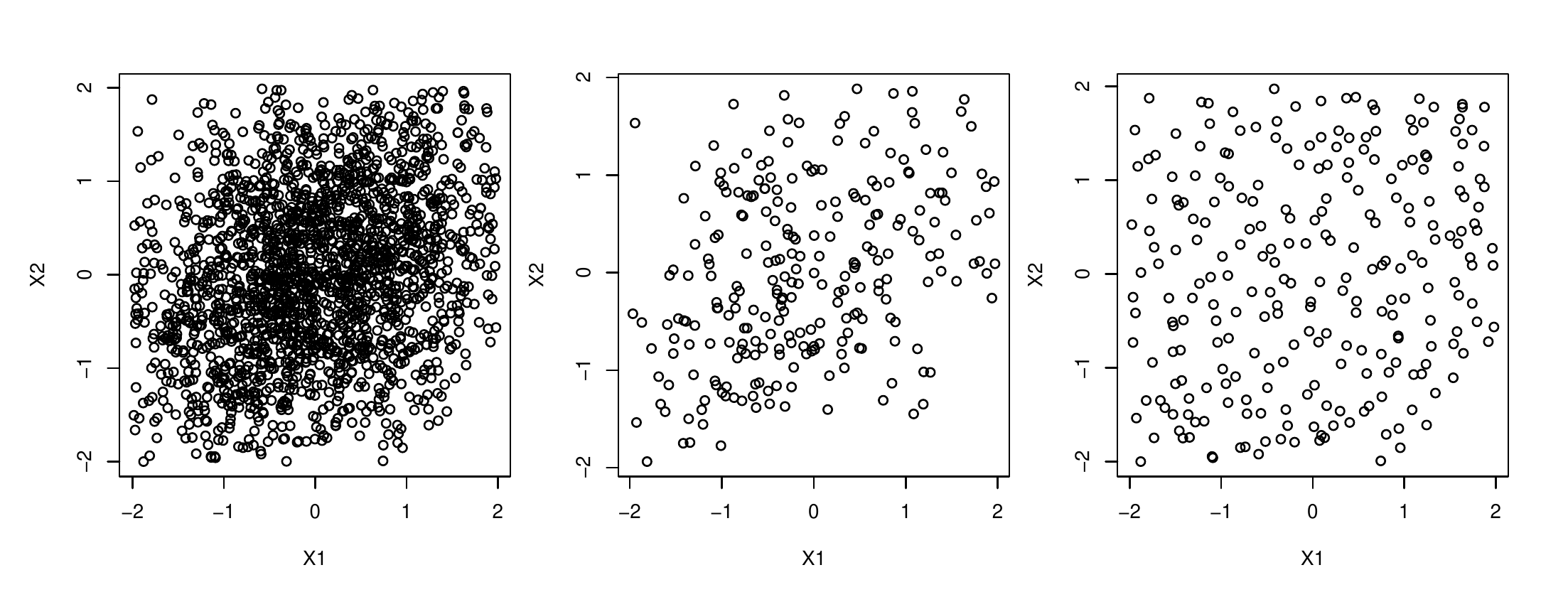}
\caption{Illustration of Algorithm \ref{alg:oss} with simulated data. The full sample (left), a random subsample (middle), and the IES subsample (right).}
 \label{fig:proj}
\end{figure}

To visualize the resulting subsample of Algorithm \ref{alg:oss}, we generate full data of $2000$ i.i.d. bivariate normal points, truncated in absolute value by 2. The generating distribution has zero mean, unit variance and a correlation of $0.3$ between any two predictors. Figure~\ref{fig:proj} plots the full data (left), a random subsample (middle), and an IES subsample (right), both subsamples of size $250$. The hyperparameter $q=16$ is used for the IES.  Figure~\ref{fig:proj} clearly shows that predictors in the IES subsample are more uniformly distributed and less correlated than predictors in the random subsample.

\section{Simulation Studies}\label{sec:simu} 

In this section, we evaluate the performance of the IES method through simulation studies. We compare the IES subsample with the random subsample (Rand) and the LowCon method. LowCon is a subsampling method developed in \citet{ma_eff_spline_spacefilling_2020} for smoothing splines and in \citet{meng_lowcon_2021} for misspecified linear models. It selects a subsample that approximates a prefix space-filling design \citep{joseph2015maximum,lin2015latin} via nearest neighbor search. 


We set the full sample size $N=10000$ and generate values of $p=3$ predictors from two distributional settings:
\begin{itemize}[leftmargin=*,label={}]
    \item{Case 1.} The predictors follow a truncated multivariate normal $\mathcal{TN}(0,\Sigma,-2,2)$ with mean zero and covariance matrix 
    $
        \Sigma=(0.3^{\mathbbm{1}{(i\neq j)}}).
    $
    Each predictor lies in $[-2,2]$.
    
    \item{Case 2.} The predictors are generated via a truncated multivariate exponential distribution using the elliptical copula in the R package \textit{copula}. The covariance matrix $\Sigma$ is the same as in Case 1. The marginal distribution is specified as an exponential with rate one, and is truncated above by $4$ and translated to $[-2,2]$.
\end{itemize}

The responses are generated by $Y=m(\mathbf{X})+\epsilon$, where
\begin{equation}\label{Meansim}
m(\mathbf{X})=1+\frac{8}{4+X_1}+\frac{\exp\left\{3-X_2^2\right\}}{4}+1.5\sin\left(\frac{\pi}{2}X_3\right),
\end{equation}
and $\epsilon$ follows $\mathcal{N}(0,0.25)$. 

The effect of model misspecification on IES is also studied, where an additional interaction term $2\ln (4.5+X_1X_2)$ is added to the true regression function in \eqref{Meansim} but is not used when training an additive model. 
 
\begin{figure}[t]
\centering
\includegraphics[scale=0.8]{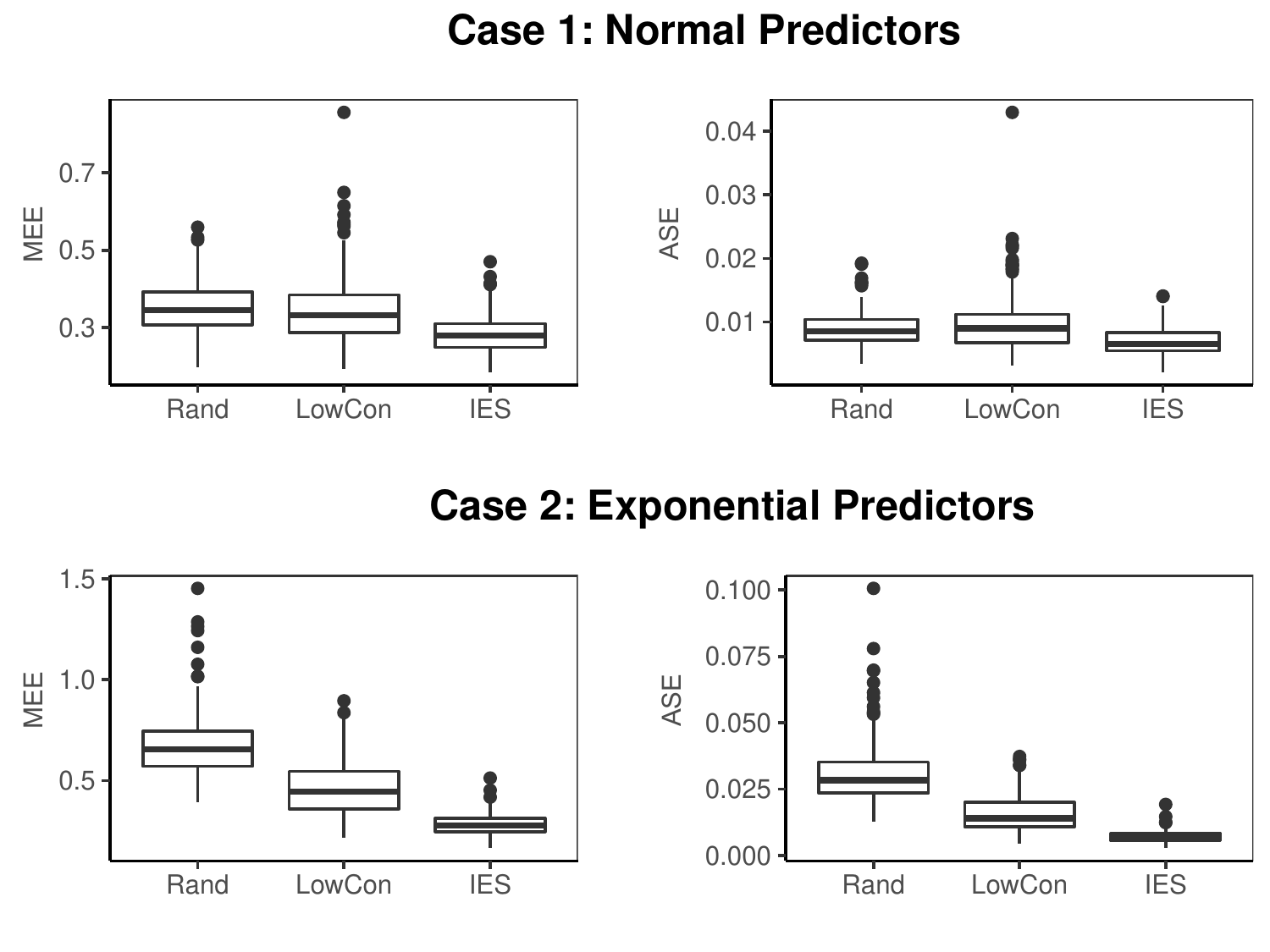}
\caption{The $\rm{MEE}$ (left) and $\rm{ASE}$ (right) of $\hat{m}$ trained on different subsamples of the full sample in the two cases.} 
\label{Perform1}
\end{figure}

\begin{figure}[t]
\centering
\includegraphics[scale=0.2]{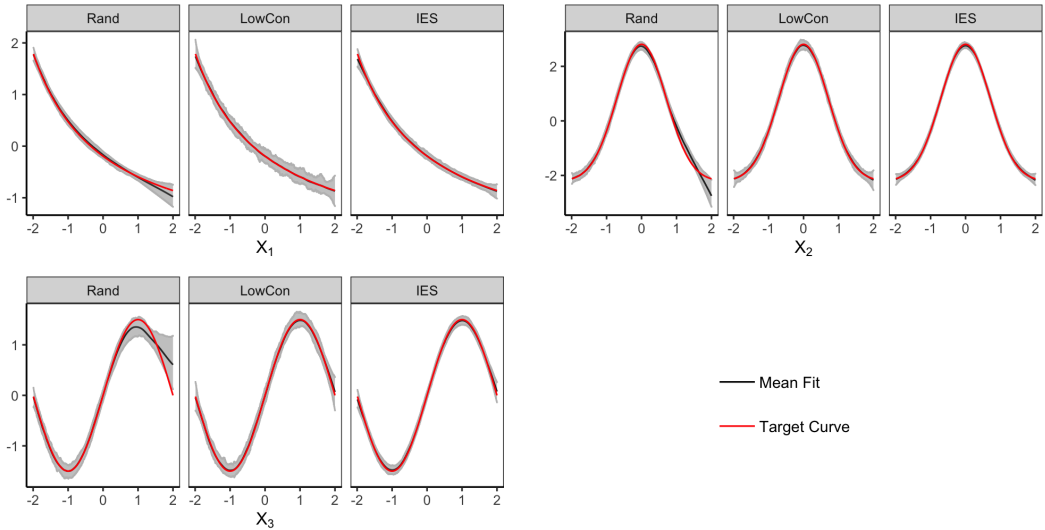}
\caption{Component function estimates trained on subsamples obtained by different methods for Case 2: exponentially distributed predictors.}
\label{Perform2}
\end{figure}

Each setting of predictors is replicated $200$ times, and the three subsampling methods, Rand, LowCon and IES, are performed for each replication with the subsample size $n=1000$. The hyperparameter $q=16$ is used for the IES method. Backfitting algorithm with local linear smoothers is then applied to train an additive model over each subsample. The bandwidth, searched in $\{0.05,0.1,0.15,\dots,0.95\}^3$, is chosen via a five-fold cross validation (CV). 
For the $\hat{m}$ trained over each subsample, we consider two performance measures, namely, the maximum estimation error
${\rm MEE}=\max_{\mathbf{x}\in\mathcal{X}_{test}}|\hat{m}(\mathbf{x})-m(\mathbf{x})|,$
and the average squared error ${\rm ASE}=\sum_{\mathbf{x}\in\mathcal{X}_{test}}\left(\hat{m}(\mathbf{x})-m(\mathbf{x})\right)^2/10^6.$
The ${\rm MEE}$ is a realization of the maximum risk used in \eqref{eq:modmm} and quantifies the worst performance of $\hat{m}$, and the ${\rm ASE}$ measures the overall performance of $\hat{m}$ over the test domain. The test data $\mathcal{X}_{test}$ are $10^{6}$ grid points with each predictor spanning at 100 evenly spaced points from $-1.8$ to $1.8$. 
 
Figure \ref{Perform1} plots the ${\rm MEE}$ and ${\rm ASE}$ of $\hat{m}$ trained on different subsamples across the 200 replications. 
The IES consistently allows better estimation of $m$ than the subsamples selected from other methods. Specifically, the MEE plots demonstrate the advantage of the IES in controlling the worst error across the entire domain, and the ASE plots suggest a better overall estimation performance of IES. 

Figure \ref{Perform2} depicts the fitted curves of each component function in \eqref{Meansim} for each subsample of the full data generated in Case 2. The red curve represents the target centered component function, and the black curve indicates the average fit over the $200$ replications. The grey shaded area is the empirical $95\%$ confidence band. It is clear that the IES method always outperforms random subsampling in allowing a better fit of each component function. When compared with LowCon, the IES performs similarly in terms of average fit, but it performs better in terms of stability (width of the shaded band), especially in the area with low density, i.e. the right tails of all component functions, and when the target function assumes a nonlinear shape, e.g. the turnings areas in the second and third component functions. Figure \ref{fig:normcompare} in Supplementary Materials reveals similar comparison results for the subsamples
of the full data generated in Case 1.

The out-performance of IES over LowCon comes from two aspects. Firstly, the IES samples diverse points sequentially and avoids duplicates, while LowCon applies nearest neighborhood search to approximate a prefix space-filling design, which often samples repeatedly on the same observation in the region with scarce data. Duplicated points have bigger weights and increase the modeling instability. Secondly, most space-filling designs target at full dimensional uniformity but may not be uniform when projected to low dimensions. IES targets at one- and two-dimensional uniformity and thus is more suitable for establishing additive models.

\begin{figure}[t]
\centering
\includegraphics[scale=0.8]{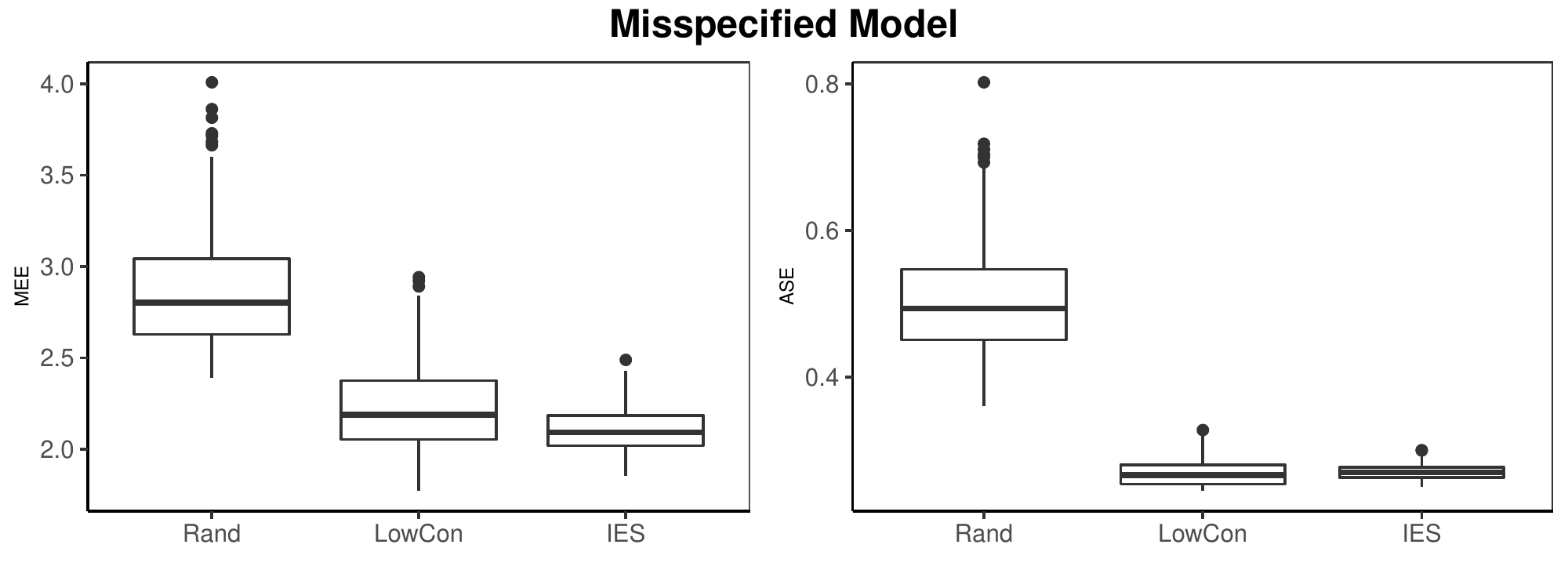}
\caption{The MEE (left) and ASE (right) on the regression function with misspecification.}
\label{mis}
\end{figure}



Figure~\ref{mis} shows the boxplots of MEEs and ASEs for the regression function with the misspecified interaction term $2\ln (4.5+X_1X_2)$. The predictors are generated the same as in Case 2. The lower estimation error for IES suggests that its subsamples are less susceptible to model misspecification because of the fact that the predictors in an IES subsample are less dependent. In our particular setting, $X_3$ is nearly independent of $X_1$ and $X_2$ in the IES subsample. Hence, the component function of $X_3$ is not affected by the misspecified interaction term of $X_1$ and $X_2$ and be accurately estimated. The plots of estimated component functions are relegated to Figure \ref{fig:miscompare} in Supplementary Materials to save space. 


\section{Real Data}\label{sec:real}

We now evaluate the performance of the IES method on the Diamond Price Prediction dataset. The dataset is available from both the R package \textsf{ggplot2} and \url{https://www.kaggle.com/shivam2503/diamonds}. 
\emph{Price} along with $9$ predictors of 53,940 diamonds are collected in the data with the goal of building a predictive model for the diamond price. Three discrete quality measures, namely \emph{cut}, \emph{color}, and \emph{clarity}, are dropped, as we focus on continuous predictors. Among continuous predictors, \emph{carat, depth} (which summarizes information in other left-out predictors) and \emph{table}, are picked for modeling. The first predictor measures the weight of each diamond and the latter two are specialized shape metrics. Since both \emph{carat} and \emph{price} are highly skewed, a log transformation is applied. We train the model $$\emph{price} \approx \mu+m_{1}(\emph{carat})+m_{2}(\emph{depth})+m_{3}(\emph{table})$$
over selected subsamples via the same backfitting procedure as in Section \ref{sec:simu}.

\subsection{Estimation Performance}

Backfitting on the LowCon subsample of this dataset does not converge. Therefore, we only compare the IES with random subsamples. We use the model trained on the full data as a benchmark because the true model is unknown to us. The subsample size is fixed at $n=5000$. 

\begin{figure}[t]
\includegraphics[scale=0.6]{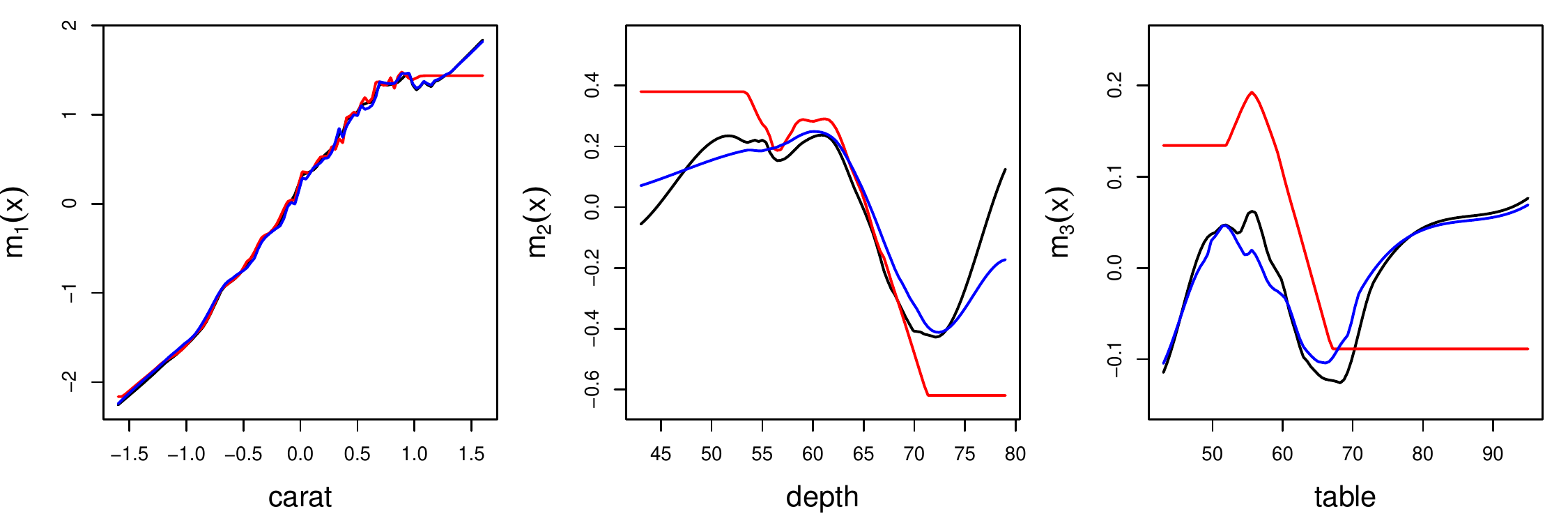}
\centering
\caption{Centered component function estimates obtained on the full data (black), random subsample (red), and IES subsample (blue). }
\label{fig:3}
\end{figure}

Figure~\ref{fig:3} depicts estimated component functions trained on the full sample and  subsamples selected by different methods. The span of $x$-axis of each component function reflects its range in the full data. Since a subsample often results in a reduced range of predictors, extrapolation is needed. In this case, we use term-wise nearest neighbor estimation. 
In Figure~\ref{fig:3}, the component function of \emph{carat} has a dominant effect in magnitude with mostly a linear shape. The estimations over an IES subsample and a random subsample are both close to the benchmark, with the IES showing its advantage in the right tail. This confirms that the IES subsample provides better worst-case control in accuracy. The estimation of the other two component functions clearly demonstrates the superiority of IES. The IES effectively captures the information of each component function, even if the function has a complex shape and a relatively weak signal. 

\begin{table}[t] \centering 
  \caption{Estimation and prediction performances of Rand and IES in the diamond price prediction data.} 
  \label{table:dia1} 
\begin{tabular}{@{\extracolsep{5pt}} cccc} 
\\[-1.8ex]\hline 
\hline \\[-1.8ex] 
 & Rand & IES \\ 
\hline \\[-1.8ex] 
$\rm ASE$  & $0.13$ & $0.01$ \\ 
$\rm MEE$  & $1.51$ & $0.48$ \\ 
AvePredError & $0.06$ & $0.06$ \\ 
MaxPredError & $1.67$ & $1.29$ \\ 
\hline \\[-1.8ex] 
\end{tabular} 
\end{table}

Table~\ref{table:dia1} further compares the performance of IES and random subsamples using measures for estimation and prediction errors. 
First, same as in Section \ref{sec:simu}, we calculate MEE and ASE for the regression function over the test data $\mathcal{X}_{test}$, the grid points of size $10^6$ that span the range of the full data. The response for $\mathcal{X}_{test}$ is generated using the model trained on the full data. In addition, we calculate the average (AvePredError) and maximum prediction error (MaxPredError) for the observed \emph{price} in the full data. From Table~\ref{table:dia1}, an IES subsample outperforms a random subsample in minimizing both estimation and prediction errors. An additive model trained on an IES subsample provides more accurate component function estimation and response prediction than the model trained over a random subsample. 

\subsection{Computation Time}

We now report the computational time of IES on the Diamond data. 
Table \ref{table:ct} lists the computation time of subsampling, CV, and model fitting procedures as well as the total spent time, with their respective standard deviations shown in parenthesis. 
As shown in Table~\ref{table:ct}, CV dominates the time consumption for training an additive model, making the modeling on the full data dramatically slow. Training the model on a subsample significantly accelerates the CV and reduces the time to around 8-fold. The IES sampling procedure does take a few more seconds, but this is unimportant compared to the big saving on the time for CV. The total time of IES and Rand are comparable, and it makes sense for IES to be a little slower than Rand to achieve its superior estimation performance.

\begin{table}[t] \centering 
  \caption{Average computation times (in seconds) spent on subsampling, CV, and model fitting. Standard deviations (SD) are in parentheses.  }
  \label{table:ct} 
\begin{tabular}{@{\extracolsep{\fill}}llllllllll}
\\[-1.8ex]
\hline 
\hline \\[-1.8ex] 
 & Full & Rand & IES \\ 
\hline \\[-1.8ex] 
Subsampling & $0$ $(0)$& $0.0003$ $(0.0000)$& $5.82$ $(0.34)$\\ 
CV & $8092.53$ $(121.99)$& $926.06$ $(20.12)$& $1140.04$ $(27.52)$\\ 
Fitting & $0.21$ $(0.11)$& $0.03$ $(0.02)$& $0.03$ $(0.01)$\\ 
Total & $8092.74$ $(121.96)$& $926.09$ $(20.12)$& $1145.89$ $(27.48)$&\\
\hline \\[-1.8ex] 
\end{tabular} 
\end{table}

\section{Discussion}
\label{sec:conc}
We have developed a new subsampling method, called ISE, to accelerate the computation of training an additive model from large data. The ISE selects the subsample that approximates an OA and optimizes the minimax risk of training an additive model by enabling asymptotically independent and uniformly distributed predictors in the selected subsample. Theoretical results have been derived to guarantee the convergence of the backfitting procedure over an ISE subsample for two-dimensional problems. 
Extensive simulation studies and a real data application demonstrate that ISE outperforms existing subsampling methods in providing accurate estimations of the regression function and each component. 

Future works can look into subsampling via OAs with higher strength. The asymptotic property in Theorem \ref{thm:asym} can be easily extended to a general number of predictors if the training subsample has a higher strength. In addition, such a subsample achieves higher-order independence among multiple predictors and will allow better estimation of an additive model with interaction terms. Another direction is to consider the performance of IES for a more general family of models, for example, the generalized additive model. We expect that such a subsample will perform well for estimating $g(E[Y])$ for a general link function $g$ because of its independence between predictors and uniform coverage of the data region.

\section*{Supplementary Materials}\label{sec:supp}

The supplementary materials include the proofs of Theorems~\ref{l1}--\ref{thm:asym} and additional simulation results. 

\begin{appendices}

\section{Existence of OA}\label{appenda}
The existence and construction of OAs have been widely studied in the literature, see, for example, \citet{hedayat1999orthogonal} and \citet{dey2009fractional} for a comprehensive introduction. Below is a well-known result.

\begin{lmm}\label{oalem1}
If $q$ is a prime power and $\lambda$ is a positive integer, then an OA$(\lambda q^2,p,q,2)$ exists for any $p\le q+1$. 
\end{lmm}
A construction of OA$(q^2,q+1,q,2)$ with $q$ being a prime power can be found in \citet{hedayat1999orthogonal} (Theorem $3.1$). Stacking $\lambda$ copies of an OA$(q^2,q+1,q,2)$ provides an OA$(\lambda q^2,q+1,q,2)$, any $p$ columns of which is an OA$(\lambda q^2,p,q,2)$. 

When $q$ is not a prime power, one may construct OAs from pairwise orthogonal Latin squares. The lemma below comes from this approach. 

\begin{lmm}\label{oalem2}
    Let $q_1^{v_1}q_2^{v_2}\cdots q_u^{v_u}$ be a prime factorization of $q$ and $q_{0}=\min\{q_i^{v_i}\mid i=1,\dots,u\}$, then an OA$(\lambda q^2,p,q,2)$ exists for any $p\leq q_0+1$.
\end{lmm}

The result is an immediate consequence of Theorems $8.4$ and $8.28$ in \citet{hedayat1999orthogonal}. It extends $q$ from prime power to an arbitrary positive integer.

Many other OAs with flexible $p$ and $q$ exist, see \url{http://neilsloane.com/oadir/} for a collection of examples.



\section{The unique solution in Theorem \ref{thm:asym}}\label{appendb}

Denote the observations in the IES subsample $\{\mathcal{X}^*_{opt},\mathbf{y}^*_{opt}\}$ by $(x^*_{i1},x^*_{i2},y^*_i)$ where $i\in\{1,2,\dots,n\}$, $x^*_{i1}$ and $x^*_{i2}$ are the two predictors, and $y^*_i$ is the response. Define, for $t=0, 1, 2$,
$$V_{nt}(x)=\frac{1}{n}\sum_{i=1}^n\frac{1}{h_1}K\left(\frac{x^*_{i1}-x}{h_1}\right)(x^*_{i1}-x)^t, \  \text{and} \  
W_{nt}(x)=\frac{1}{n}\sum_{i=1}^n\frac{1}{h_2}K\left(\frac{x^*_{i2}-x}{h_2}\right)(x^*_{i2}-x)^t.
$$

Then define $n\times n$ matrices $\mathcal{S}_1=\{[\mathcal{S}_1]_{ij}\}_{1\leq i, j \leq n}$ and $\mathcal{S}_2=\{[\mathcal{S}_2]_{ij}\}_{1\leq i, j \leq n}$ where
\renewcommand{\theequation}{\thesection\arabic{equation}}
\begin{align}\label{eq:s1}
   [\mathcal{S}_1]_{ij}&=\frac{\frac{1}{nh_1}K\left(\frac{x^*_{j1}-x^*_{i1}}{h_1}\right)V_{n2}(x^*_{i1})-\frac{1}{nh_1}K\left(\frac{x^*_{j1}-x^*_{i1}}{h_1}\right)(x^*_{j1}-x^*_{i1})V_{n1}(x^*_{j1}) }{V_{n0}(x^*_{i1})V_{n2}(x^*_{i1})-V_{n1}(x^*_{i1})^2},\\
   \text{and} \quad [\mathcal{S}_2]_{ij}&=\frac{\frac{1}{nh_2}K\left(\frac{x^*_{j2}-x^*_{i2}}{h_2}\right)W_{n2}(x^*_{i2})-\frac{1}{nh_2}K\left(\frac{x^*_{j2}-x^*_{i2}}{h_2}\right)(x^*_{j2}-x^*_{i2})W_{n1}(x^*_{j2}) }{W_{n0}(x^*_{i2})W_{n2}(x^*_{i2})-W_{n1}(x^*_{i2})^2}.\nonumber
\end{align}
Following \citet{buja1989linear} and
\citet{opsomer_fitting_1997}, the bivariate additive model, fitted by local linear smoothers via backfitting algorithm, aims to solve the following estimation equation: 

\begin{equation}\label{estieq}
    \begin{pmatrix}
\mathcal{I} &\mathcal{S}_1^*\\
\mathcal{S}_2^* &\mathcal{I}\\
\end{pmatrix}
\begin{pmatrix}
\mathbf{\hat{m}}_1\\
\mathbf{\hat{m}}_2\\
\end{pmatrix}
=\begin{pmatrix}
\mathcal{S}_1^*\\
\mathcal{S}_2^*\\
\end{pmatrix}
\mathbf{Y},
\end{equation}
where $\mathbf{\hat{m}}_1=\left(\hat{m}_1(x^*_{11}),\dots,\hat{m}_1(x^*_{n1})\right)^\top$, $\mathbf{\hat{m}}_2=\left(\hat{m}_2(x^*_{12}),\dots,\hat{m}_2(x^*_{n2})\right)^\top$, $\mathbf{Y}=\left(y^*_1,\dots,y^*_n\right)^\top$, $\mathcal{S}_1^*=(\mathcal{I}-\mathbf{1}\mathbf{1}^\top/n)\mathcal{S}_1$, and 
$\mathcal{S}_2^*=(\mathcal{I}-\mathbf{1}\mathbf{1}^\top/n)\mathcal{S}_2$ with $\mathcal{I}$ being the $n \times n$ identity matrix and $\mathbf{1}$ being a $n \times 1$ vector of all ones. The centering constant $\mu$ is estimated separately by $\hat{\mu}=\bar{y}$. The backfitting algorithm on the IES subsample converges to the unique solution 
\begin{equation*}
   \begin{pmatrix}
\mathbf{\hat{m}}_1\\
\mathbf{\hat{m}}_2\\
\end{pmatrix} 
=\begin{pmatrix}
\left[\mathcal{I}-\left(\mathcal{I}-\mathcal{S}_1^*\mathcal{S}_2^*\right)^{-1}\left(\mathcal{I}-\mathcal{S}_1^*\right)\right]\mathbf{Y}\\
\left[\mathcal{I}-\left(\mathcal{I}-\mathcal{S}_2^*\mathcal{S}_1^*\right)^{-1}\left(\mathcal{I}-\mathcal{S}_2^*\right)\right]\mathbf{Y}\\
\end{pmatrix}.
\end{equation*}

\end{appendices}

\bibliographystyle{chicago}
\bibliography{Bography}
\makeatletter\@input{withOSS.tex}\makeatother
\end{document}


\def\spacingset#1{\renewcommand{\baselinestretch}%
{#1}\small\normalsize} \spacingset{1}

\if1\blind
{
   \title{\bf
 Supplementary Materials for ``Independence-Encouraging Subsampling for 
  Nonparametric Additive Models"
  }
  
  \maketitle
} \fi

\if0\blind
{
  \bigskip
  \bigskip
  \bigskip
  \begin{center}
    {\LARGE\bf Title}
\end{center}
  \medskip
} \fi

The document contains the proofs of Theorems~\ref{l1}--\ref{thm:asym} and additional simulation results. 

\spacingset{1.45}
\section{Proofs}\label{sec:proofs}

\subsection{
\textbf{Proof of Theorem~\ref{l1}}}
\begin{proof}
 For any fixed $x$ in the support, define $m_0(\cdot)=\left(b_N/2\right)[1-\sqrt{\eta}(\cdot-x)^2/b_N]_{+}$, where $[a]_{+}=\max\{0,a\}$ and $b_N=\left[15\eta^{1/4}\sigma^2/\left(Nf(x)\right)\right]^{2/5}$. Then $m_0\in \mathcal{C}^*$. By Eq. $A.3$ of \citet{fan_desada_1992}, 
  \begin{equation}\label{s1}
  \frac{3}{4}15^{-1/5}\eta^{1/5}\left(\frac{\sigma^2}{N}\right)^{4/5} f(x)^{-4/5}(1+o_p(1))\le E[(\tilde{m}(x)-m_0(x))^2|\mathbf{X}_1,\dots,\mathbf{X}_N],
  \end{equation}
for any linear smoother $\tilde{m}$.
Fix $x$ at $x_0=\arg\min_{x\in [0,1]} f(x)$ on the left side of \eqref{s1}, and it follows from the definition of $\sup_{m\in \mathcal{C}^*,x\in[0,1]}E[(\tilde{m}(x)-m(x))^2|\mathbf{X}_1,\dots,\mathbf{X}_N]$ that
$$\frac{3}{4}15^{-1/5}\eta^{1/5}\left(\frac{\sigma^2}{N}\right)^{4/5} f(x_0)^{-4/5}(1+o_p(1))\le \sup_{m\in \mathcal{C}^*,x\in[0,1]}E[(\tilde{m}(x)-m(x))^2|\mathbf{X}_1,\dots,\mathbf{X}_N].$$
Thus 
\begin{equation}\label{lowbd}
    R(f)\ge\frac{3}{4}15^{-1/5}\eta^{1/5}\left(\frac{\sigma^2}{N}\right)^{4/5} f(x_0)^{-4/5}(1+o_p(1)).
\end{equation}

It suffices to show that the lower bound in \eqref{lowbd} is also an upper bound for $R(f)$. 
Consider the local linear estimator $\hat{m}_L$ using the kernel $K_0(u)=\frac{3}{4}(1-u^2)_{+}$
and bandwidth $h_0=\left(\frac{15\sigma^2}{f(x)\eta N}\right)^{1/5}.$
Evaluating the MSE with this particular linear smoother $\hat{m}_L$ and taking the supremum gives that
$$\sup_{m\in \mathcal{C}^*,x\in[0,1]}E[(\hat{m}_L(x)-m(x))^2|\mathbf{X}_1,\dots,\mathbf{X}_N]=\frac{3}{4}15^{-1/5}\eta^{1/5}\left(\frac{\sigma^2}{N}\right)^{4/5} \min_{x\in[0,1]}f(x)^{-4/5}(1+o_p(1)).$$
This completes the proof.
\end{proof}

\subsection{
\textbf{Proof of Theorem~\ref{oalemma}}}


We provide and prove a more general result, Lemma S\ref{subsampwoa} below, and Theorem \ref{oalemma} will follow as a special case of Lemma S\ref{subsampwoa}. 
We need the concept of weak strength \citep{Xu_2003}.
\begin{defn}
    An $n\times p$ design with $q$ levels is called an OA of weak strength $t$, denoted as OA$(n,p,q,t^{-})$, if all level combinations for any $t$ columns appear as equally often as possible, that is, the difference of occurrence of level combinations does not exceed one in any $t$ columns.
\end{defn}


\begin{lemma}\label{subsampwoa}
For a subsample $\mathcal{X}^*$, 
\begin{equation}\label{lxstar}
L(\mathcal{X}^*)\ge\frac{p(p-1)h(n,q^2)+ph(n,q)-np^2}{2},
\end{equation} 
where $h(a,b)=\floor{a/b}^2b+\left(2\floor{a/b}+1\right)\left(a-\floor{a/b}b\right)$. The lower bound in \eqref{lxstar} is achieved if and only if the membership matrix of $\mathcal{X}^*$ is an OA$(n,p,q,t^{-})$ for $t=1,2$.
\end{lemma}

\begin{proof}
Let $\mathcal{Z}^*$ be the membership matrix of $\mathcal{X}^*$ and define $$K(\mathcal{Z}^*)=\frac{2}{n(n-1)}\sum_{1\leq i< i'\leq n}\lbrack\delta(\mathbf{z}_i^*,\mathbf{z}_{i'}^*)\rbrack^2,$$
where $\mathbf{z}_{i}^*$ and $\mathbf{z}_{i'}^*$ are two disctinct rows in $\mathcal{Z}^*$.
By Lemma 1 and Corollary $3(ii)$ of \citet{Xu_2003}, we have $$K(\mathcal{Z})\ge\frac{p(p-1)h(n,q^2)+ph(n,q)-np^2}{n(n-1)}$$
for any $\mathcal{Z}\in \{0,1,\dots,q-1\}^{n\times p}$, and the
equality holds if and only if $Z$ is an OA$(n,p,q,t^-)$ for $t=1,2$. Hence,
$$L(\mathcal{X}^*)=\frac{n(n-1)}{2}K(\mathcal{Z}^*)\ge\frac{p(p-1)h(n,q^2)+ph(n,q)-np^2}{2}.$$
This completes the proof.
\end{proof}


\begin{proof}[Proof of Theorem \ref{oalemma}]
An OA of strength 2 is of both weak strength $1^-$ and $2^-$, so Lemma S\ref{subsampwoa} applies. Take $n$ to be a multiple of $q^2$. Then $h(n,q^2)=n^2/q^2$ and $h(n,q)=n^2/q$. Substitution of both expressions into Equation \eqref{lxstar} completes the proof.
\end{proof}



\subsection{
\textbf{Proof of Theorem~\ref{thm:inp}}}


The following lemma is needed to prove Theorem \ref{thm:inp}.

\begin{lemma}\label{primeoa}
Given that an OA$(q^2,p+1,q,2)$ exists, an OA$(n,p,q,2^-)$ that is simultaneous of strength $1^-$ exists for any positive integer $n$. 
\end{lemma}

\begin{proof}
We prove the lemma by construction. Let $\lambda=[n/q^2]$, where $[\cdot]$ denotes the closest integer. Then $-q^2\leq n-\lambda q^2\leq q^2$. We consider the case $g(q)=n-\lambda q^2<0$. The case $g(q)>0$ follows the same construction by adding a copy of the selected rows to the OA$(\lambda q^2,p,q,2)$ constructed below.

Start with an OA$(q^2,p+1,q,2)$. Arrange its rows so that the first column is ascending in levels, from $0$'s to $(q-1)$'s, and denote this OA by $\mathcal{A}$. Stacking $\lambda$ copies of it forms an OA$(\lambda q^2,p+1,q,2)$, denoted as $\mathcal{A}'$. Denote the submatrix consisting of the first $|g(q)|$ rows from $\mathcal{A}$ by $\mathcal{A}_c$ and
delete the first column of $\mathcal{A}_c$. Then $\mathcal{A}_c$ is an OA$(|g(q)|,p,q,t^-)$ for $t=1,2$. Delete the first column and the first $|g(q)|$ rows in $\mathcal{A}'$.
The resulting matrix, as a complement of $\mathcal{A}_c'$, is then an OA$(n,p,q,t^{-})$ for $t=1,2$. The result is thus proved. 
\end{proof}

We now prove Theorem \ref{thm:inp} under the following weaker assumption in replace of Assumption \ref{a1}. 
\begin{assumption}\label{as1}
    $n-\lambda q^2=O(q^{\gamma})$ for some fixed positive integer $\lambda$ and $\gamma\in(0,2)$, and an OA$(q^2, p+1, q, 2)$ exists. 
\end{assumption}

\begin{proof}[Proof of Theorem~\ref{thm:inp}]
The proof consists of two parts: 
\begin{itemize}
\item [(i).] For any integer $\lambda>0$, with probability approaching one, a full data, under Assumptions \ref{a2}--\ref{a3}, covers all $q^p$ cells that constitute the $p$ dimensional unit cube at least $\lambda+1$ times. 
\item [(ii).]
An IES subsample is sufficiently close to a random OA under the Assumptions~\ref{a1} and \ref{a3}.
We prove for the case $\lambda=1$, since the proof for $\lambda\neq 1$ is essentially the same.  Without loss of generality, we assume that all $p$ predictors take values in $[0,1]$.
\end{itemize}

\medskip

\noindent 
\emph{Proof of (i)}. 
Let $\left(\mathbf{X}_1,\dots,\mathbf{X}_N\right)^{T}$ denote a random predictor matrix of dimension $N\times p$ satisfying Assumption~\ref{a2}. Define $E_N$ as the event that $\left(\mathbf{X}_1,\dots,\mathbf{X}_N\right)^{T}$ occupies all $q^p$ cells at least 
$\lambda+1$ times. Denote $B_l$ the event that the $l$-th cell is occupied at most $\lambda$ times. By Assumption~\ref{a2}, there exists $a\in(0,1)$ and $b>a$ such that the joint density of predictors is larger than $a$ and smaller than $b$. When $\lambda=1$, $$P(B_l)<\left(1-\frac{a}{q^p}\right)^{N}+N\frac{b}{q^p}\left(1-\frac{a}{q^p}\right)^{N-1},$$ for all $l \in\{1,\dots,q^p\}.$
It follows that \[P(E_{N}^C)=P\left(\bigcup_{l=1}^{q^p} B_l\right)\le \sum P(B_l) <q^p\left(1-\frac{a}{q^p}\right)^{N}+Nb\left(1-\frac{a}{q^p}\right)^{N-1}.\numberthis \label{upb}\] 
The two terms in the upper bound in \eqref{upb} can be rewritten as 
\begin{align*}
q^p\left(1-\frac{a}{q^p}\right)^{N}&=\exp\left\{\ln{q^p}+N\ln\left(1-\frac{a}{q^p}\right)\right\}\\
&=\exp\left\{\ln{q^p}-Na/q^p+o(N/q^{2p})\right\},\numberthis \label{eqn}
\end{align*}
and
\begin{align*}
Nb\left(1-\frac{a}{q^p}\right)^{N-1}&=b\exp\left\{\ln{N}+(N-1)\ln\left(1-\frac{a}{q^p}\right)\right\}\\
&=b\exp\left\{\ln{N}-(N-1)a/q^p+o(N/q^{2p})\right\}.\numberthis \label{eqn2}
\end{align*}
Equations \eqref{eqn} and \eqref{eqn2} come from Taylor expansion of $\ln(1-a/q^p)$, where $q$ goes to infinity as implied by Assumptions~\ref{a1} and \ref{a3}. Under the two assumptions, $q=O(N^{\nu/2})$ for some $\nu\in(0,2/p)$ and the first term in Equations \eqref{eqn} and \eqref{eqn2} is of order $O(\log N)$. Therefore, the second term, of order $\Omega(N^{1-\nu p/2})$, dominates the first. As a result, both equations goes to $\exp\{-\infty\}=0$. This proves $\lim_{N\rightarrow\infty}P(E_N)=1$ and concludes the first part.





\medskip

\noindent 
\emph{Proof of (ii)}. 
By Lemma S\ref{subsampwoa}, the minimizer of $L$ has their membership matrix as an OA$(n,p,q,t^{-})$ and for $t=1,2$ if such an OA exists. The existence is then guaranteed by Assumption S\ref{as1} and Lemma S\ref{primeoa}. Moreover, in view of Part (i), the probability that the full data contains a subsample with such an OA membership matrix approaches one as $N\rightarrow\infty$. $\mathcal{X}^*_{opt}$ is thus guaranteed to have its membership matrix as a OA$(n,p,q,2^{-})$ in probability. 
By Assumption S\ref{as1} and with $\lambda=1$, $n=q^2+g(q)$ for some $g(q)=O(q^\gamma)$. According to the definition of OA$(q^2+g(q),p,q,2^{-})$, the membership matrix of $\mathcal{X}_{opt}^*$ is different from an OA$(q^2,p,q,2)$ by $|g(q)|$ rows on any two columns. 

For any $x_1,x_2\in[0,1]$, the quantity $\sum_{i=1}^{n}\mathbbm{1}(X^*_{ij}\le x_1, X^*_{ij'}\le x_2)$ is bounded between $\floor{x_1q}\floor{x_2q}$ and $(\floor{x_1q}+1)\left(\floor{x_2q}+1\right)$ for a subsample with an OA$(q^2,p,q,2)$ membership matrix. Hence with probability approaching one, induced distribution of $\mathcal{X}^*_{opt}$ satisfies
$$\frac{(\floor{x_1q})\left(\floor{x_2q}\right)-|g(q)|}{q^2+g(q)}\le
F_n(x_1,x_2)\le \frac{(\floor{x_1q}+1)\left(\floor{x_2q}+1\right)+|g(q)|}{q^2+g(q)}.$$ Then, in probability,
$$\left|F_n(x_1,x_2)-x_1x_2\right|\leq \frac{2q+2|g(q)|+1}{q^2+g(q)}.$$
By Assumptions S\ref{as1} and \ref{a3}, the above bound is of order $O(q^{-\min\{1,2-\gamma\}})=O(N^{-\min\{1,2-\gamma\}\nu/2})$, and does not depend on $x_1$ and $x_2$. 
Together with Part (i), 
\begin{align*}
 \sup_{x_1,x_2}\left|F_n(x_1,x_2)-x_1x_2\right|=O_p(N^{-\min\{1,2-\gamma\}\nu/2}).
\end{align*}
Setting $\gamma=1$ gives the result in Theorem \ref{thm:inp}.

\end{proof}


\subsection{
\textbf{Proof of Theorem~\ref{thm:asym}}}

We first generalize the standard definition of Lipschitz functions and 
 introduce the definition of $M_n$-Lipschitz functions to ease the
presentation of the proof. 

\begin{defn}[$M_n$-Lipschitz function]\label{defn:lips}
We call a sequence of functions $g_n$ defined on $\mathcal{T}$ as $M_n$-Lipschitz if for any $s, t \in \mathcal{T}$,
$$
|g_n(s) - g_n(t)| \leq M_n |s-t|.
$$  
\end{defn}

We now introduce some notations. 

\begin{itemize}[leftmargin=0.5cm]
\item Throughout the proofs below, constants are absolute, that is, they do not vary with $n$, $q$ or $h.$
    \item 
Denote $O_{UP}$ as a big $O_P$ term that is uniform in $x\in[0,1]$. Formally, 
we write 
$T_n(x)=O_{UP}(\alpha_n)$ 
if $\sup_{x\in[0,1]}|T_n(x)|=O_P(\alpha_n)$.

Similarly we use $o_{UP}$ to denote the uniform $o_P$ counterpart. 


\item Given a point $x\in[0,1]$ and a positive bandwidth $h$, define
  $\mathcal{D}_{x,h}=\{u\in[-1,1]:x+uh\in[0,1]\}$. A point $x \in [0, 1]$
  is called an interior point if $\mathcal{D}_{x,h}=[-1,1]$; otherwise, it is called a boundary
  point. 

  \item
  Define the $t$-th 
  boundary moment as  $$
R_t(x;h):=\int_{\mathcal{D}_{x,h}}K(u)u^tdu,\quad t=0,1,2.
$$ 
Denote $M_0=\sup_{u\in[-1,1]}|K(u)|$. Then we have \begin{equation}\label{RtLip}
    |R_t(x;h)-R_t(x';h)|\le\left|\int_{\mathcal{D}_{x,h}-\mathcal{D}_{x',h}}M_0du\right|\le \frac{2M_0}{h}|x-x'|,
\end{equation}
where $\mathcal{D}_{x,h}-\mathcal{D}_{x',h}$ is the symmetric difference between the two sets. 
This shows that $R_t(\cdot;h)$ is $(2M_0/h)$-Lipschitz.


\item Recall that $E_N$ is the event that the full data covers each of the $q^2$ grids at least $\lambda+1$ times. By (i) in the proof of 
Theorem \ref{thm:inp}, $P(E_N) \rightarrow 1$  as the full sample size $N \rightarrow \infty$. 

\end{itemize}


Next we provide some technical lemmas that will be used to prove Theorem~\ref{thm:asym}. Note that ``$n\rightarrow \infty$'' in all the proofs below can be implied by ``$N\rightarrow \infty$'' due to Assumption \ref{a3}. Moreover, $qh^2=O(\sqrt{n})h^2$ by Assumption \ref{a1} and $1/(qh^2)$ goes to $0$ by Assumption \ref{a5}.


Recall that $V_{nt}(x)$ and $W_{nt}(x)$, $t=0,1,2,$ are defined in Appendix 
B. We provide a lemma to show that each of them is essentially a Riemann sum and its corresponding approximation error can be properly controlled.

\begin{lemma}\label{VWappr}
    Under Assumptions \ref{a2}-\ref{a5} and conditioning on the event $E_N$,
    \begin{equation*}
      V_{nt}(x)=h_1^tR_t(x;h_1)+        \delta_t,  
\ \text{and} \  W_{nt}(x)=h_2^tR_t(x;h_2)+\zeta_t, 
      \ t=0,1,2.
    \end{equation*}
where $\delta_t$ and $\zeta_t$ are diminishing error terms such that $\left|\delta_t/h_1^t\right|=O_{UP}(1/(qh_1^2))$ and $\left|\zeta_t/h_2^t\right|=O_{UP}(1/(qh_2^2))$ .
Here we inhibit the dependence on $x$ in the notations of $\delta_t$ and $\zeta_t$ for simplicity.

\end{lemma}

\begin{proof}

We consider two cases to prove the result for $V_{nt}(x)$. 

\noindent \textbf{Case 1: $n-\lambda q^2=0$.} 
When $n-\lambda q^2=0$ and $E_N$ occurs, $\mathcal{X}^*_{opt}$ has an OA
membership matrix by Theorem \ref{oalemma}. Therefore, $\mathcal{X}^*_{opt}$
covers each of the $q$ by $q$ grids $\lambda$ times. This suggests that
each $V_{nt}, t=0, 1, 2$, is essentially a Riemann sum as shown in details below.

For each $k\in\{1,\dots,q\}$, 
let $\{x_{k,s}\}_{s\in\{1,2,\dots,\lambda q\}}$
denote a subset of $\{x^*_{i1}: i=1, \ldots, n\}$ that falls into $[(k-1)/q,k/q]$. 
Since $n=\lambda q^2$, we have  
\begin{align}
     V_{nt}(x)&=
     \frac{h_1^t}{\lambda q}\sum_{i=1}^{\lambda q^2}\frac{1}{h_1}K\left(\frac{x^*_{i1}-x}{h_1}\right)\left(\frac{x^*_{i1}-x}{h_1}\right)^{t}\frac{1}{q}
     \nonumber\\
     &=\frac{h_1^t}{\lambda q}\sum_{s=1}^{\lambda q}\sum_{k=1}^{q}\frac{1}{h_1}K\left(\frac{x_{k,s}-x}{h_1}\right)\left(\frac{x^*_{k,s}-x}{h_1}\right)^{t}\frac{1}{q}\label{eq:sumbase}\\
     &=\frac{h_1^t}{\lambda q}\sum_{s=1}^{\lambda q}\left(\int_{0}^{1}\frac{1}{h_1}K\left(\frac{x'_{s}-x}{h_1}\right)\left(\frac{x'_{s}-x}{h_1}\right)^{t}dx'_{s}+\delta_{t,s}\right)\nonumber\\
     &=h_1^t\int_{0}^{1}\frac{1}{h_1}K\left(\frac{x'-x}{h_1}\right)\left(\frac{x'-x}{h_1}\right)^{t}dx'+\delta_t\nonumber\\
     &=h_1^t\int_{\mathcal{D}_{x,h_1}}K\left(u\right)u^tdu+\delta_t\nonumber,
\end{align}
where the last step is obtained by letting $u=(x'-x)/h_1$, 
$$\delta_{t,s}=\sum_{k=1}^{q}\frac{1}{h_1}K\left(\frac{x_{k,s}-x}{h_1}\right)\left(\frac{x_{k,s}-x}{h_1}\right)^{t}\frac{1}{q}-\int_{0}^{1}\frac{1}{h_1}K\left(\frac{x'_{s}-x}{h_1}\right)\left(\frac{x'_{s}-x}{h_1}\right)^{t}dx'_{s},$$ 
and 
$\delta_t=\{h_1^t/(\lambda q)\}\sum_{s=1}^{\lambda q}\delta_{t,s}$. Here we inhibit the dependence on $x$ in the notations of $\delta_{t,s}$ and $\delta_t$ for simplicity.


We next provide a bound for $\delta_{t,s}$ which is uniform in $x\in[0,1]$. By
the mean value theorem, there is a set of real numbers $\{x^{(k)}\in[(k-1)/q,k/q]: k=1,2,\dots,q\}$ such that 
\begin{equation}\label{mvthm}
\int_{(k-1)/q}^{k/q}\frac{1}{h_1}K\left(\frac{x'_{s}-x}{h_1}\right)\left(\frac{x'_{s}-x}{h_1}\right)^{t}dx'_{s}=\frac{1}{h_1}K\left(\frac{x^{(k)}-x}{h_1}\right)\left(\frac{x^{(k)}-x}{h_1}\right)^{t}\frac{1}{q}.
\end{equation}
For $t=0,1,2$ and all $s$,
\begin{align}\label{deltats}
    |\delta_{t,s}|&=\left|\sum_{k=1}^{q}\frac{1}{qh_1}\left\{K\left(\frac{x_{k,s}-x}{h_1}\right)\left(\frac{x_{k,s}-x}{h_1}\right)^{t}-K\left(\frac{x^{(k)}-x}{h_1}\right)\left(\frac{x^{(k)}-x}{h_1}\right)^{t}\right\}\right|\nonumber\\
    &\le \frac{1}{h_1}\max_{k}\left|K\left(\frac{x_{k,s}-x}{h_1}\right)\left(\frac{x_{k,s}-x}{h_1}\right)^{t}-K\left(\frac{x^{(k)}-x}{h_1}\right)\left(\frac{x^{(k)}-x}{h_1}\right)^{t}\right|\nonumber\\
    &\le \frac{1}{h_1}\max_{k}\left[\left|K\left(\frac{x_{k,s}-x}{h_1}\right)\left\{\left(\frac{x_{k,s}-x}{h_1}\right)^{t}-\left(\frac{x^{(k)}-x}{h_1}\right)^{t}\right\}\right|\right.\nonumber\\
&\indent\left.+\left|\left\{K\left(\frac{x_{k,s}-x}{h_1}\right)-K\left(\frac{x^{(k)}-x}{h_1}\right)\right\}\left(\frac{x^{(k)}-x}{h_1}\right)^{t}\right|\right] \nonumber\\
&\le\frac{1}{h_1}\max_{k}\left\{\left|K\left(\frac{x_{k,s}-x}{h_1}\right)\frac{x_{k,s}-x^{(k)}}{h_1}B_{t,k,s}\right|+M\left|\frac{x_{k,s}-x^{(k)}}{h_1}\right|\right\},
\end{align}
where $B_{0,k,s}=0$, 
$$B_{t,k,s}=\sum_{l=0}^{t-1}\left(\frac{x_{k,s}-x}{h_1}\right)^{l}\left(\frac{x^{(k)}-x}{h_1}\right)^{t-1-l}, \quad t=1, 2,$$ 
and the second term in \eqref{deltats} is obtained via Assumption \ref{a4}. 


We next discuss bounds of $B_{t,k,s}$ and thus those of $|\delta_{t,s}|$ via \eqref{deltats}.

\begin{itemize}[leftmargin=0.5cm]
\item
When $t=0$, we have $B_{0,k,s}=0$. Accompanied by the fact that, for all $k$ and $s$, $|x_{k,s}-x^{(k)}|\le 1/q$,
we have $|\delta_{0,s}|\le M/(qh_1^2)$.

\item
When $t=1$, we have $B_{1,k,s}=1$. Hence $|\delta_{1,s}|\le (M_0+M)/(qh_1^2)$, where $M_0=\sup_{u\in[-1,1]}|K(u)|$.

\item
When $t=2$ and $x_{k,s}\not\in\supp{K((\cdot-x)/h_1)}$, we have $K((x_{k,s}-x)/h_1)(x_{k,s}-x^{(k)})/h_1B_{2,k,s}=0$, which implies that $|\delta_{2,s}|\le M/(qh_1^2)$.

\item 
When $t=2$ and $x_{k,s}\in\supp{K((\cdot-x)/h_1)}$, we have $((k-1)/q,k/q)\cap \supp{K((\cdot-x)/h_1)}\neq\emptyset$ almost surely. The left hand side of \eqref{mvthm} is positive, so $x^{(k)}\in\supp{K((\cdot-x)/h_1)}\subset[x-h_1,x+h_1]$. As a result, 
$|B_{2,k,s}|=\left|(x_{k,s}-x)/h_1+(x^{(k)}-x)/h_1\right|\le 2,$
and $|\delta_{2,s}|\le (2M_0+M)/(qh_1^2)$.
\end{itemize}

In summary,  $|\delta_{t,s}|\le (tM_0+M)/(qh_1^2)$ for $t=0,1,2$. Therefore, 
$$\left|\frac{\delta_t}{h_1^t}\right| \le \frac{1}{\lambda q} \sum_{s=1}^{\lambda q} |\delta_{t,s}|\le \frac{tM_0+M}{qh_1^2} = O_{UP}\left(
    \frac{1}{qh_1^2}\right).
$$

\noindent \textbf{Case 2: $n-\lambda q^2\neq 0$.} Under Assumption \ref{a1}, $g(q)=n-\lambda q^2=O(q)$. When $E_N$ occurs, the membership matrix of $\mathcal{X}^*_{opt}$ has $|g(q)|$ rows different from an OA$(\lambda q^2,2,q,2)$. Without loss of generality, we assume $g(q)>0$ and the first $\lambda q^2$ rows in $\mathcal{X}^*_{opt}$ forms an OA. Then
\begin{align}\label{Vncase2}
        V_{nt}(x)&=
     \frac{h_1^t}{\lambda q+g(q)/q}\sum_{i=1}^{\lambda q^2}\frac{1}{h_1}K\left(\frac{x^*_{i1}-x}{h_1}\right)\left(\frac{x^*_{i1}-x}{h_1}\right)^{t}\frac{1}{q}\nonumber\\
     &\indent+\frac{h_1^t}{\lambda q+g(q)/q}\sum_{i=\lambda q^2+1}^{\lambda q^2+g(q)}\frac{1}{h_1}K\left(\frac{x^*_{i1}-x}{h_1}\right)\left(\frac{x^*_{i1}-x}{h_1}\right)^{t}\frac{1}{q}.
\end{align}
By \textbf{Case 1}, the integral approximation to the first term of \eqref{Vncase2} gives an error term of $O_{UP}(h_1^{t-2}/q)$.
Since $g(q)=O(q)$, and $\left|1/h_1K\left((x'-x)/h_1\right)\left((x'-x)/h_1\right)^{t}\right|\le M_0/h_1$ for any $x,x'\in[0,1]$,
the second term in \eqref{Vncase2} is 
of order $O_{UP}(h_1^{t-1}/q)$, which is $o_{UP}(h_1^{t-2}/q)$.
Thus the result remains the same as in \textbf{Case 1}.




The result for $W_{nt}(x)$ can be proved following similar arguments, so we omit its proof.
\end{proof}

We next provide bounds for 
\begin{align*}
Q_t(x;h):=\frac{R_t(x;h)}{R_0(x;h)R_2(x;h)-R_1(x;h)^2}, \quad t=1, 2.
\end{align*}
\begin{lemma}\label{Q}
 Under Assumption \ref{a4},
$$\sup_{h\in(0,1/2]}\sup_{x\in[0,1]}|Q_1(x;h)|\le\frac{4\int_{0}^{1}K(u)du}{\int_{0}^{1}2K(u)u^2du-\left(\int_{0}^{1}2K(u)udu\right)^2},$$
\begin{equation*}
     \inf_{h\in(0,1/2]}\inf_{x\in[0,1]}Q_2(x;h)\ge \frac{1}{2},\text{ and } \sup_{h\in(0,1/2]}\sup_{x\in[0,1]}Q_2(x;h)\le\frac{4\int_{0}^{1}2K(u)u^2du}{\int_{0}^{1}2K(u)u^2du-\left(\int_{0}^{1}2K(u)udu\right)^2}.
\end{equation*}
Particularly, $Q_1(x;h)=0$ and $Q_2(x;h)=1$ for each interior point $x$.
\end{lemma}
\begin{proof}
For any fixed $h\in(0,1/2]$, if $x$ is an interior point, then $R_0(x;h)=1$, $R_1(x;h)=0$, and $R_2(x;h)=\int_{0}^{1}2K(u)u^2du$. Hence $Q_1(x;h)=0$ and $Q_2(x;h)=1$.

For arbitrary $x\in[0,1]$ and $h\in(0,1/2]$, exploiting the fact that kernel $K$ is a symmetric density on $[-1,1]$, we have  
$1/2\le R_0(x;h)\le1$, $0\le|R_1(x;h)|\le\int_{0}^{1}K(u)du$, and $\int_{0}^{1}K(u)u^2du\le R_2(x;h)\le\int_{0}^{1}2K(u)u^2du$.  
Along with the Cauchy-Schwartz inequality,
\begin{align}\label{detbound}
R_0(x;h)R_2(x;h)-R_1^2(x;h)\ge\frac{1}{4}\left\{\int_{0}^{1}2K(u)u^2du-\left(\int_{0}^{1}2K(u)udu\right)^2
\right\}>0.
\end{align}
From the bounds for $R_t(x;h)$ and \eqref{detbound}, it follows
$$|Q_1(x;h)|\le\frac{4\int_{0}^{1}K(u)du}{\int_{0}^{1}2K(u)u^2du-\left(\int_{0}^{1}2K(u)udu\right)^2},$$ and $$\frac{1}{2}\le Q_2(x;h)\le\frac{4\int_{0}^{1}2K(u)u^2du}{\int_{0}^{1}2K(u)u^2du-\left(\int_{0}^{1}2K(u)udu\right)^2}.$$
Note that the bounds for $Q_1$ and $Q_2$ above depend on neither $x$ nor $h$.  The proof is thus complete.

\end{proof}

Define $f_1(r_0,r_1,r_2)=r_1/(r_0r_2-r_1^2)$ and $f_2(r_0,r_1,r_2)=r_2/(r_0r_2-r_1^2)$. 
The next lemma reveals the relation between $\left(R_0(x;h),R_1(x;h),R_2(x;h)\right)$ and $Q_t(x;h)$ via $f_t, t=1,2$, and then establishes the Lipschitz continuity of each $Q_t(x;h)$ accordingly.




\begin{lemma}\label{f_t}
    Let $\eta_0$, $\eta_1$ and $\eta_2$ be three terms of order $o_{UP}\left(1\right)$ and $h\in(0,1/2]$. Under Assumption \ref{a4}, for $t=1,2$,
    \begin{align} \label{fqrelation}       f_t\left(R_0(x;h)+\eta_0,R_1(x;h)+\eta_1,R_2(x;h)+\eta_2\right)=Q_t(x;h)\left\{1+O_{UP}\left(\eta_0+\eta_1+\eta_2\right)\right\}.
    \end{align} Furthermore, for any $h\in(0,1/2]$ and $t\in\{1,2\}$, $Q_t(\cdot;h)=f_t\left(R_0(\cdot;h),R_1(\cdot;h),R_2(\cdot;h)\right)$ is $(C_t/h)$-Lipschitz for some constant $C_{t}>0$. 
\end{lemma}
\begin{proof}
Clearly, for any $\epsilon>0$, $f_t(r_0,r_1,r_2)$ is twice continuously
differentiable on $\{(r_0,r_1,r_2):r_0r_2-r_1^2\ge \epsilon\}$. Hence, by \eqref{detbound}, $f_t(r_0,r_1,r_2)$ is twice continuously differentiable on $$\mathcal{T}:=[1/2,1]\times\left[-\int_{0}^{1}K(u)du,\int_{0}^{1}K(u)udu\right]\times\left[\int_{0}^{1}K(u)u^2du,\int_{0}^{1}2K(u)u^2du\right],$$ which is a compact set 
that 
contains $\{(R_0(x;h),R_1(x;h),R_2(x;h)):x\in[0,1]\}$ for $h\in(0,1/2]$.  We now apply the Taylor expansion at $\left(R_0(x;h),R_1(x;h),R_2(x;h)\right)$. For  each $t=1,2$,
\begin{align*}
&f_t\left(R_0(x;h)+\eta_0,R_1(x;h)+\eta_1,R_2(x;h)+\eta_2\right)\\
&=Q_t(x;h)+\frac{\partial f_t}{\partial r_0}\left(R_0(x;h),R_1(x;h),R_2(x;h)\right)\eta_0+\frac{\partial f_t}{\partial r_1}\left(R_0(x;h),R_1(x;h),R_2(x;h)\right)\eta_1\\
&\quad+\frac{\partial f_t}{\partial r_2}\left(R_0(x;h),R_1(x;h),R_2(x;h)\right)\eta_2+o_{UP}\left(\eta_0+\eta_1+\eta_2\right),
\end{align*}
where the $o_{UP}$ term is due to the boundedness of the second-order partial derivatives of $f_t$ on $\mathcal{T}$. 

Let $D_{t',t}=\sup_{(r_0,r_1,r_2)\in\mathcal{T}}|\partial f_t/\partial r_{t'}(r_0,r_1,r_2)|$ with $t'=0,1,2$ and $t=1,2$. By the continuity of $\partial f_t/\partial r_{t'}$ and compactness of $\mathcal{T}$, each $D_{t',t}$ is a bounded constant. Along with $\eta_{t'}=o_{UP}(1)$ for $t'=0,1,2$, we have 
$$
\frac{\partial f_t}{\partial r_{t'}}\left(R_0(x;h),R_1(x;h),R_2(x;h)\right)\eta_{t'}=O_{UP}(\eta_{t'}), \quad t'=0,1,2. 
$$
By Lemma S\ref{Q}, $\sup_{h\in(0,1/2]}\sup_{x\in[0,1]}|Q_t(x;h)|$ is bounded from above by some constant. Hence 
$$Q_t(x;h)+O_{UP}\left(\eta_0+\eta_1+\eta_2\right)=Q_t(x;h)\left\{1+O_{UP}\left(\eta_0+\eta_1+\eta_2\right)\right\}.$$
This proves \eqref{fqrelation}.

We next show that $Q_t(\cdot;h)$ is $(C_t/h)$-Lipschitz. Let $x,x'\in[0,1]$. By the multi-variable mean value theorem and continuity of $R_0,R_1,R_2$, there exist $\nu_0,\nu_1,\nu_2\in[0,1]$ such that
\begin{align}\label{multimeanvt}
&\indent f_t\left(R_0(x';h),R_1(x';h),R_2(x';h)\right)-f_t\left(R_0(x;h),R_1(x;h),R_2(x;h)\right)\nonumber\\
&=\frac{\partial f_t}{\partial r_0}\left(R_0(\nu_0;h),R_1(\nu_1;h),R_2(\nu_2;h)\right)\left(R_0(x';h)-R_0(x;h)\right)\nonumber\\
&\indent+\frac{\partial f_t}{\partial r_1}\left(R_0(\nu_0;h),R_1(\nu_1;h),R_2(\nu_2;h)\right)\left(R_1(x';h)-R_1(x;h)\right)\nonumber\\
&\indent+\frac{\partial f_t}{\partial r_2}\left(R_0(\nu_0;h),R_1(\nu_1;h),R_2(\nu_2;h)\right)\left(R_2(x';h)-R_2(x;h)\right).
\end{align}
Recall that $D_{t',t}=\sup_{(r_0,r_1,r_2)\in\mathcal{T}}|\partial f_t/\partial r_{t'}(r_0,r_1,r_2)|$ with $t'=0,1,2$ and $t=1,2$.
 Combined with \eqref{multimeanvt}, we have 
\begin{align*}
    &\indent \left|f_t\left(R_0(x';h),R_1(x';h),R_2(x';h)\right)-f_t\left(R_0(x;h),R_1(x;h),R_2(x;h)\right)\right|\nonumber\\
&\le D_{0,t}\left|\left(R_0(x';h)-R_0(x;h)\right)\right|+D_{1,t}\left|\left(R_1(x';h)-R_1(x;h)\right)\right|\nonumber\\
&\indent+D_{2,t}\left|\left(R_2(x';h)-R_2(x;h)\right)\right|\nonumber\\
&\le \left(D_{0,t}+D_{1,t}+D_{2,t}\right)\frac{2M_0}{h}|x-x'|.&& \{\text{by \eqref{RtLip}}\}
\end{align*}
Taking $C_t=2M_0\left(D_{0,t}+D_{1,t}+D_{2,t}\right)$ completes the proof.
\end{proof}

The next lemma shows that $Q_t, t=1, 2$ convoluted with the kernel
is also Lipschitz. 

\begin{lemma}\label{intlip}
   Under Assumption \ref{a4}, there exists a contant $\kappa_t$ for all $h\in(0,1/2]$ such that the integrals $(1/h)\int_{0}^{1}K\left((x-\nu)/h\right)Q_t(\nu;h)d\nu$ and
   $(1/h)\int_{0}^{1}K\left((x-\nu)/h\right)\left((x-\nu)/h\right)Q_t(\nu;h)d\nu$, as functions of $x$ on $[0,1]$, are $(\kappa_t/h)$-Lipschitz, $t=1,2$.
\end{lemma}
\begin{proof}
    We first consider $(1/h)\int_{0}^{1}K\left((x-\nu)/h\right)Q_t(\nu;h)d\nu$. Recall that, by Lemma S\ref{Q}, $Q(\cdot;h)$ is $(C_t/h)$-Lipschitz, and $\sup_{h\in(0,1/2]}\sup_{x\in[0,1]}|Q_t(x;h)|$ is bounded above by some constant, say, $\xi_t>0$. 
    Let $x,x'\in[0,1]$ and set $u=(\nu-x)/h$ and $u'=(\nu-x')/h$. Then
    \begin{align}
       &\indent\frac{1}{h}\left|\int_{0}^{1}K\left(\frac{x-\nu}{h}\right)Q_t(\nu;h)d\nu-\int_{0}^{1}K\left(\frac{x'-\nu}{h}\right)Q_t(\nu;h)d\nu\right|\nonumber\\
    &=\left|\int_{\mathcal{D}_{x,h}}K(u)Q_t(x+uh;h)du-\int_{\mathcal{D}_{x',h}}K(u')Q_t(x'+u'h;h)du'\right|  \nonumber\\
    &\le\left|\int_{\mathcal{D}_{x,h}\cap\mathcal{D}_{x',h}}K(u)\left\{Q_t(x+uh;h)-Q_t(x'+uh;h)\right\}du\right| \nonumber\\
    &\indent+\left|\int_{\mathcal{D}_{x,h}-\mathcal{D}_{x',h}}\sup_{u\in[0,1]}|K(u)|\sup_{x\in[0,1]}|Q_t(x;h)| du\right| \nonumber\\
    &\le\left|\int_{\mathcal{D}_{x,h}\cap\mathcal{D}_{x',h}}M_0\frac{C_t}{h}|x-x'|du\right|+\left|\int_{\mathcal{D}_{x,h}-\mathcal{D}_{x',h}}M_0\xi_tdu\right|&&\{\text{by Lemma S\ref{Q}}\}\nonumber\\
    &\le\frac{2M_0(C_t+\xi_t)}{h}|x-x'|\nonumber,
    \end{align}
where $\mathcal{D}_{x,h}-\mathcal{D}_{x',h}$ is the symmetric difference between the two sets.
\noindent
Similarly, 
\begin{align*}
       &\indent\frac{1}{h}\left|\int_{0}^{1}K\left(\frac{x-\nu}{h}\right)\frac{x-\nu}{h}Q_t(\nu;h)d\nu-\int_{0}^{1}K\left(\frac{x'-\nu}{h}\right)\frac{x'-\nu}{h}Q_t(\nu;h)d\nu\right|\nonumber\\
       &\indent\le\frac{2M_0(C_t+\xi_t)}{h}|x-x'|.
\end{align*}
The desired result is proved with $\kappa_t=2M_0(C_t+\xi_t)$.
\end{proof}

Recall that $(x^*_{i1},x^*_{i2})$ is the $i$-th row of $\mathcal{X}^*_{opt}$, $i\in\{1,2,\dots,n\}$.

\begin{lemma}\label{1dRieSum}
    Let $\{g(x;h)\}_{h\in(0,1/2]}$ be a family of functions defined on $[0,1]$ such that $g(\cdot;h)$ is $(C/h^{\alpha})$-Lipschitz, for some constant $C>0$ and $\alpha\le 1$. If there exists a constant $G>0$ such that $\sup_{h\in(0,1/2]}\sup_{x\in[0,1]}|g(x;h)| \leq G$, then conditioning on $E_N$ and under Assumptions \ref{a2}-\ref{a5}, for $t\in\{0,1\}$ and $\gamma\in\{1,2\}$, we have   \begin{align}\label{summand2}
        &\frac{1}{n}\sum_{i=1}^{n}K\left(\frac{x^*_{il}-x_0}{h_{\gamma}}\right)\left(\frac{x^*_{il}-x_0}{h_{\gamma}}\right)^{t}g(x^{*}_{i\gamma};h_{\gamma})\nonumber\\
        &=\int_{0}^{1}K\left(\frac{x-x_0}{h_{\gamma}}\right)\left(\frac{x-x_0}{h_{\gamma}}\right)^{t}g(x;h_{\gamma})dx+O_{UP}\left(\frac{1}{qh_{\gamma}}\right). 
    \end{align}
\end{lemma}
\begin{proof}
Here we only prove the result for $\gamma=1$, since the proof for $\gamma=2$ is exactly the same.
We consider a large enough $n$ so that $h_1\le 1/2$.

We first establish the Lipschitz continuity of $K\left((x^*_{il}-x_0)/h_{1}\right)\left((x^*_{il}-x_0)/h_{1}\right)^{t}g(x^{*}_{i1};h_{1})$ 
as a function of $x^{*}_{i1}$ for the case $t=0 \text{ and }1$. 
Note that $g(\cdot;h_1)$ is $(C/h_1^{\alpha})$-Lipschitz on $[0,1]$, and, by Assumption \ref{a4}, $K((\cdot-x_0)/h_1)$ is $(M/h_1)$-Lipschitz on $[0,1]$. 
Therefore, 

\begin{itemize}[leftmargin=0.5cm]
\item $t=0$.
\begin{align*}  &\indent\left|K\left(\frac{x-x_0}{h_1}\right)g(x;h_1)-K\left(\frac{x'-x_0}{h_1}\right)g(x';h_1)\right|\\ 
&\le \left|\left\{K\left(\frac{x-x_0}{h_1}\right)-K\left(\frac{x'-x_0}{h_1}\right)\right\}g(x;h_1)\right|+\left|K\left(\frac{x'-x_0}{h_1}\right)\left\{g(x;h_1)-g(x';h_1)\right\}\right|\\
&
\le \left(\frac{M\sup_{x}|g(x;h_1)|}{h_1}+\frac{CM_0}{h_1^{\alpha}}\right)|x-x'|\le\frac{MG+CM_0}{h_1}|x-x'|, 
\end{align*}
which shows that $K((\cdot-x_0)/h_1)((\cdot-x_0)/h_1)^tg(\cdot;h_1)$ is
$\left\{(MG+CM_0)/h_1\right\}$-Lipschitz. 


\item $t=1$. Since $K((x-x_0)/h)=K((x'-x_0)/h)=0$ whenever both $x\not\in[x_0-h_1,x_0+h_1]$ and $x'\not\in[x_0-h_1,x_0+h_1]$, it suffices to only consider the case where
at least one of $x$ and $x'$ falls into $[x_0-h_1,x_0+h_1]$. 

If $x'\in[x_0-h_1,x_0+h_1]$, then  without loss of generality
\begin{align}\label{prodlipsch2}
&\indent\left|K\left(\frac{x-x_0}{h_1}\right)g(x;h_1)\left(\frac{x-x_0}{h_1}\right)-K\left(\frac{x'-x_0}{h_1}\right)g(x';h_1)\left(\frac{x'-x_0}{h_1}\right)\right|\nonumber\\
&\le \left|K\left(\frac{x-x_0}{h_1}\right)g(x;h_1)\right|\left|\frac{x-x_0}{h_1}-\frac{x'-x_0}{h_1}\right|\nonumber\\
&+\left|K\left(\frac{x-x_0}{h_1}\right)g(x;h_1)-K\left(\frac{x'-x_0}{h_1}\right)g(x';h_1)\right|\left|\left(\frac{x'-x_0}{h_1}\right)\right|\nonumber\\
&\le \frac{M_0G}{h_1}|x-x'|+\frac{MG+CM_0}{h_1}|x-x'|=\frac{C'}{h_1}|x-x'|,
\end{align}
where $C':=M_0 G + MG + C M_0$, so $K((\cdot-x_0)/h_1)((\cdot-x_0)/h_1)^tg(\cdot;h_1)$ 
is $(C'/h_1)$-Lipschitz for $t=1$.



\end{itemize}


Next we follow similar arguments as in the proof of Lemma S\ref{VWappr} to prove \eqref{summand2}. Explicitly,  

\begin{itemize}[leftmargin=0.5cm]
\item $n=\lambda q^2$. We have  
\begin{align*}
    |\tau_{t,s}|&=\left|\sum_{k=1}^{q}\frac{1}{q}\left\{K\left(\frac{x_{k,s}-x_0}{h_1}\right)\left(\frac{x_{k,s}-x_0}{h_1}\right)^{t}g(x_{k,s};h_1)-K\left(\frac{x^{(k)}-x_0}{h_1}\right)\left(\frac{x^{(k)}-x_0}{h_1}\right)^{t}g\left(x^{(k)};h_1\right)\right\}\right|\\
    &\le\max_{k}\left|K\left(\frac{x_{k,s}-x_0}{h_1}\right)\left(\frac{x_{k,s}-x_0}{h_1}\right)^{t}g(x_{k,s};h_1)-K\left(\frac{x^{(k)}-x_0}{h_1}\right)\left(\frac{x^{(k)}-x_0}{h_1}\right)^{t}g\left(x^{(k)};h_1\right)\right|\\
    &\le\frac{C'}{h_1}\max_{k}\left|x_{k,s}-x^{(k)}\right|=\frac{C'}{qh_1},
\end{align*}
where $x_{k,s}$ and $x^{(k)}$ are defined the same as in the proof of Lemma S\ref{VWappr}, and
$$\tau_{t,s}:=\sum_{k=1}^{q}K\left(\frac{x_{k,s}-x_0}{h_1}\right)\left(\frac{x_{k,s}-x_0}{h_1}\right)^{t}g(x_{k,s};h_1)\frac{1}{q}-\int_{0}^{1}K\left(\frac{x_{s}-x_0}{h_1}\right)\left(\frac{x_{s}-x_0}{h_1}\right)^{t}g(x_{s};h_1)dx_{s}.$$ 
Therefore the integral approximation error  $\tau_t:=1/(\lambda q)\sum_{s=1}^{\lambda q}\delta_{t,s}=O_{UP}\left(1/(qh_1)\right)$. 

\item $n\neq\lambda q^2$. The proof is also similar to that of Lemma S\ref{VWappr} and is omitted. 

\end{itemize}

\end{proof}

\begin{lemma}\label{2driemannintegral}
    Let $\{g_1(x;h)\}_{h\in(0,1/2]}$ and $\{g_2(x;h)\}_{h\in(0,1/2]}$ be two families of continuous functions on $[0,1]$. Suppose that, for all $h\in(0,1/2]$, $g_2(\cdot;h)$ is $(C/h^{\alpha})$-Lipschitz with constant $C>0$ and $\alpha\le 1$. 
    Moreover, there exists a constant $G>0$ such that $\sup_{h}\sup_{x}|g_1(x;h)| \leq G$ and $\sup_{h}\sup_{x}|g_2(x;h)|\leq G$. 
    Then conditioning on $E_N$ and under Assumptions \ref{a2}-\ref{a5}, for  $t_1,t_2\in\{0,1\}$ and any $i,j$, we have
    \begin{align*}
        RS_{ij}(g_1,g_2;t_1,t_2)&:=\frac{1}{n}\sum_{l=1}^{n}\left\{K\left(\frac{x^*_{l1}-x^*_{i1}}{h_1}\right)\left(\frac{x^*_{l1}-x^*_{i1}}{h_1}\right)^{t_1}g_1(x^*_{i1};h_1)\right.\\ 
       &\indent\left.\times K\left(\frac{x^*_{j2}-x^*_{l2}}{h_2}\right)\left(\frac{x^*_{j2}-x^*_{l2}}{h_2}\right)^{t_2}g_2(x^*_{l2};h_2)\right\}\\        
&=\left\{\int_{0}^{1}K\left(\frac{x-x^*_{i1}}{h_1}\right)\left(\frac{x-x^*_{i1}}{h_1}\right)^{t_1}g_1(x^{*}_{i1};h_1)dx\right\}\\
&\quad \times
\left\{\int_{0}^{1}K\left(\frac{x^*_{j2}-x'}{h_2}\right)\left(\frac{x^*_{j2}-x'}{h_2}\right)^{t_2}g_2(x';h_2)dx'\right\}+O_{UP}\left(\frac{1}{qh_1}+\frac{1}{qh_2}\right).
    \end{align*}
\end{lemma}
\begin{proof}
We consider $n$ large enough so that $h_1,h_2\le 1/2$.

\medskip
\noindent    \textbf{Case 1:} $n=\lambda q^2$.   When $E_N$ occurs, $RS_{ij}(g_1,g_2;t_1,t_2)$ is the average of $\lambda$ two-dimensional Riemann sums over the same $q$ by $q$ grids. Therefore
    \begin{align*}
        RS_{ij}(g_1,g_2;t_1,t_2)&=\frac{1}{\lambda}\sum_{l=1}^{n}K\left(\frac{x^*_{l1}-x^*_{i1}}{h_1}\right)\left(\frac{x^*_{l1}-x^*_{i1}}{h_1}\right)^{t_1}g_1(x^*_{i1};h_1)\\
        &\indent\times K\left(\frac{x^*_{j2}-x^*_{l2}}{h_2}\right)\left(\frac{x^*_{j2}-x^*_{l2}}{h_2}\right)^{t_2}g_2(x^*_{l2};h_2)\frac{1}{q^2}\\
        &=\int_{[0,1]^2}K\left(\frac{x-x^*_{i1}}{h_1}\right)\left(\frac{x-x^*_{i1}}{h_1}\right)^{t_1}g_1(x^{*}_{i1};h_1)\\
        &\indent\times K\left(\frac{x^*_{j2}-x'}{h_2}\right)\left(\frac{x^*_{j2}-x'}{h_2}\right)^{t_2}g_2(x';h_2)dxdx'+\delta_{t_1t_2}^{ij}\\
        &=\int_{0}^{1}K\left(\frac{x-x^*_{i1}}{h_1}\right)\left(\frac{x-x^*_{i1}}{h_1}\right)^{t_1}g_1(x^{*}_{i1};h_1)dx\\
        &\indent\times\int_{0}^{1}K\left(\frac{x^*_{j2}-x'}{h_2}\right)\left(\frac{x^*_{j2}-x'}{h_2}\right)^{t_2}g_2(x';h_2)dx'+\delta_{t_1t_2}^{ij},
    \end{align*}
where \begin{align*}
\delta_{t_1t_2}^{ij}&=\frac{1}{\lambda}\sum_{l=1}^{n}K\left(\frac{x^*_{l1}-x^*_{i1}}{h_1}\right)\left(\frac{x^*_{l1}-x^*_{i1}}{h_1}\right)^{t_1}g_1(x^*_{i1};h_1)K\left(\frac{x^*_{j2}-x^*_{l2}}{h_2}\right)\left(\frac{x^*_{j2}-x^*_{l2}}{h_2}\right)^{t_2}g_2(x^*_{l2};h_2)\frac{1}{q^2}\\
&\indent-\int_{[0,1]^2}K\left(\frac{x-x^*_{i1}}{h_1}\right)\left(\frac{x-x^*_{i1}}{h_1}\right)^{t_1}g_1(x^{*}_{i1};h_1)K\left(\frac{x^*_{j2}-x'}{h_2}\right)\left(\frac{x^*_{j2}-x'}{h_2}\right)^{t_2}g_2(x';h_2)dxdx'.
\end{align*}
Let $\{(k_{m,1},k_{m,2}):m=1,\dots,q^2\}$ be an enumeration of $\{1,\dots,q\}^2$, and $\{(x_{m,1,s},x_{m,2,s}):m=1,\dots,q^2,s=1,\dots,\lambda\}$ denote the subset of $\{(x^*_{l1},x^*_{l2}):l=1,\dots,n\}$ that falls into $[(k_{m,1}-1)/q,k_{m,1}/q]\times[(k_{m,2}-1)/q,k_{m,2}/q]$. By the mean value theorem, we can find $\{(x^{(m)}_{1},x^{(m)}_{2})\in[(k_{m,1}-1)/q,k_{m,1}/q]\times[(k_{m,2}-1)/q,k_{m,2}/q]:m=1,\dots q^2\}$ such that 
\begin{align*}
&\int_{(k_{m,2}-1)/q}^{k_{m,2}/q}\int_{(k_{m,1}-1)/q}^{k_{m,1}/q}K\left(\frac{x-x^*_{i1}}{h_1}\right)\left(\frac{x-x^*_{i1}}{h_1}\right)^{t_1}g_1(x^{*}_{i1};h_1)K\left(\frac{x^*_{j2}-x'}{h_2}\right)\left(\frac{x^*_{j2}-x'}{h_2}\right)^{t_2}g_2(x';h_2)dxdx'\\
&=K\left(\frac{x_{1}^{(m)}-x^*_{i1}}{h_1}\right)\left(\frac{x_{1}^{(m)}-x^*_{i1}}{h_1}\right)^{t_1}g_1\left(x^{*}_{i1};h_1\right)K\left(\frac{x^*_{j2}-x_{2}^{(m)}}{h_2}\right)\left(\frac{x^*_{j2}-x_{2}^{(m)}}{h_2}\right)^{t_2}g_2\left(x_{2}^{(m)};h_2\right)\frac{1}{q^2}.
\end{align*}
Therefore
\begin{align}
&|\delta_{t_1t_2}^{ij}|\nonumber\\
    &\le\max_{m}\left|\frac{1}{\lambda}\sum_{s=1}^{\lambda}\left\{K\left(\frac{x^*_{m,1,s}-x^*_{i1}}{h_1}\right)\left(\frac{x^*_{m,1,s}-x^*_{i1}}{h_1}\right)^{t_1}g_1(x^*_{i1};h_1)\right.\right.\nonumber\\
    &\indent\left.\left.\times K\left(\frac{x^*_{j2}-x^*_{m,2,s}}{h_2}\right)\left(\frac{x^*_{j2}-x^*_{m,2,s}}{h_2}\right)^{t_2}g_2(x^*_{m,2,s};h_2)\right.\right\}\nonumber\\
    &\indent\left.-K\left(\frac{x_{1}^{(m)}-x^*_{i1}}{h_1}\right)\left(\frac{x_{1}^{(m)}-x^*_{i1}}{h_1}\right)^{t_1}g_1\left(x^{*}_{i1};h_1\right)K\left(\frac{x^*_{j2}-x_{2}^{(m)}}{h_2}\right)\left(\frac{x^*_{j2}-x_{2}^{(m)}}{h_2}\right)^{t_2}g_2\left(x_{2}^{(m)};h_2\right)\right|\nonumber\\
    &\le\max_{m}\max_{s}\left|K\left(\frac{x^*_{m,1,s}-x^*_{i1}}{h_1}\right)\left(\frac{x^*_{m,1,s}-x^*_{i1}}{h_1}\right)^{t_1}g_1(x^*_{i1};h_1)\right.\nonumber\\
    &\indent\left.\times K\left(\frac{x^*_{j2}-x^*_{m,2,s}}{h_2}\right)\left(\frac{x^*_{j2}-x^*_{m,2,s}}{h_2}\right)^{t_2}g_2(x^*_{m,2,s};h_2)\right.\nonumber\\
    &\indent\left.-K\left(\frac{x_{1}^{(m)}-x^*_{i1}}{h_1}\right)\left(\frac{x_{1}^{(m)}-x^*_{i1}}{h_1}\right)^{t_1}g_1\left(x^{*}_{i1};h_1\right)K\left(\frac{x^*_{j2}-x_{2}^{(m)}}{h_2}\right)\left(\frac{x^*_{j2}-x_{2}^{(m)}}{h_2}\right)^{t_2}g_2\left(x_{2}^{(m)};h_2\right)\right|\nonumber\\
    &\le\max_{m}\max_{s}\left[\left|K\left(\frac{x^*_{m,1,s}-x^*_{i1}}{h_1}\right)\left(\frac{x^*_{m,1,s}-x^*_{i1}}{h_1}\right)^{t_1}g_1(x^*_{i1};h_1)\right|\times\right.\nonumber\\
    &\indent \left\{\left|g_2(x^*_{m,2,s};h_2)\right|\left|K\left(\frac{x^*_{j2}-x^*_{m,2,s}}{h_2}\right)\left(\frac{x^*_{j2}-x^*_{m,2,s}}{h_2}\right)^{t_2}-K\left(\frac{x^*_{j2}-x_{2}^{(m)}}{h_2}\right)\left(\frac{x^*_{j2}-x_{2}^{(m)}}{h_2}\right)^{t_2}\right|\right.\nonumber\\
    &\indent+\left.\left|g_2(x^*_{m,2,s};h_2)-g_2\left(x_{2}^{(m)};h_2\right)\right|\left|K\left(\frac{x^*_{j2}-x_{2}^{(m)}}{h_2}\right)\left(\frac{x^*_{j2}-x_{2}^{(m)}}{h_2}\right)^{t_2}\right|\right\}\nonumber\\
    &\indent+\left|K\left(\frac{x^*_{j2}-x_{2}^{(m)}}{h_2}\right)\left(\frac{x^*_{j2}-x_{2}^{(m)}}{h_2}\right)^{t_2}g_2\left(x_{2}^{(m)};h_2\right)g_1\left(x^{*}_{i1};h_1\right)\times\right.\nonumber\\
    &\indent\left.\left.\left\{K\left(\frac{x^*_{m,1,s}-x^*_{i1}}{h_1}\right)\left(\frac{x^*_{m,1,s}-x^*_{i1}}{h_1}\right)^{t_1}-K\left(\frac{x_{1}^{(m)}-x^*_{i1}}{h_1}\right)\left(\frac{x_{1}^{(m)}-x^*_{i1}}{h_1}\right)^{t_1}\right\}\right|\right].\label{eq:kt}
\end{align}
Note that $K((\cdot-x_0)/h)((\cdot-x_0)/h)^{t}$ is $(C_t/h)$-Lipschitz for some constant $C_t>0$, $t=0,1$, indicated by the proof of Lemma S\ref{1dRieSum}.
Hence 
\begin{align*}
    |\delta_{t_1t_2}^{ij}| &\le M_0G\left(G\frac{C_{t_2}}{qh_2}+\frac{C}{qh_2^{\alpha}}M_0\right)
    +M_0G^2\frac{C_{t_1}}{qh_1},
\end{align*}
where $t_1,t_2\in \{0,1\}$. Since $\alpha \leq 1$, this indicates that $\delta_{t_1t_2}^{ij}=O_{UP}(1/(qh_{1})+1/(qh_{2}))$ for $t_1,t_2\in\{0,1\}.$

\medskip
\noindent  
\textbf{Case 2:} $n-\lambda q^2\neq 0$. By similar arguments as in the proof of Lemma S\ref{VWappr}, the result also holds.
\end{proof}

\begin{proof}[Proof of Theorem~\ref{thm:asym}]



By \citet{buja1989linear} 
and
\citet{opsomer_fitting_1997},
to prove that there exists a unique solution to \eqref{estieq} and the bivariate backfitting procedure converges to this solution with probability approaching one, it suffices to prove that 
$\limsup\limits_{n\rightarrow\infty}\|\mathcal{S}^*_1\mathcal{S}^*_2\|_{\infty}<1$ and 
$\limsup\limits_{n\rightarrow\infty}\|\mathcal{S}^*_2\mathcal{S}^*_1\|_{\infty}<1$ 
with probability approaching one, where $\|A\|_\infty=\max_{i}\sum_{j=1}^{n}|A_{ij}|$ is the maximum row sum of a matrix $A$.

Hereafter we only show
$\limsup\limits_{n\rightarrow\infty}\|\mathcal{S}^*_1\mathcal{S}^*_2\|_{\infty}<1$ 
with probability approaching one. The proof to show  $\limsup\limits_{n\rightarrow\infty}\|\mathcal{S}^*_2\mathcal{S}^*_1\|_{\infty}<1$ with probability approaching one is similar and thus omitted.




\medskip
Conditioning on $E_N$, it follows from \eqref{estieq} and Lemma S\ref{VWappr} that
\begin{align}\label{s1ij}
    [\mathcal{S}_1]_{ij}&=\frac{1}{nh_1}K\left(\frac{x^*_{j1}-x^*_{i1}}{h_1}\right)\left[\frac{R_2(x^*_{i1};h_1)+\delta_2/h_1^2}{\{R_0(x^*_{i1};h_1)+\delta_0\}\{R_2(x^*_{i1};h_1)+\delta_2/h_1^2\}-\{R_1(x^*_{i1};h_1)+\delta_1/h_1\}^2}\right.\nonumber\\
    &\left.\indent-\frac{x^*_{j1}-x^*_{i1}}{h_1}\frac{R_1(x^*_{i1};h_1)+\delta_1/h_1}{\{R_0(x^*_{i1};h_1)+\delta_0\}\{R_2(x^*_{i1};h_1)+\delta_2/h_1^2\}-\{R_1(x^*_{i1};h_1)+\delta_1/h_1\}^2}\right]\nonumber\\
    &=\frac{1}{nh_1}K\left(\frac{x^*_{j1}-x^*_{i1}}{h_1}\right)f_2\left(R_0(x^*_{i1};h_1)+\delta_0,R_1(x^*_{i1};h_1)+\frac{\delta_1}{h_1},R_2(x^*_{i1};h_1)+\frac{\delta_2}{h_1^2}\right)\nonumber\\
   &\indent-\frac{1}{nh_1}K\left(\frac{x^*_{j1}-x^*_{i1}}{h_1}\right)\frac{x^*_{j1}-x^*_{i1}}{h_1}f_1\left(R_0(x^*_{i1};h_1)+\delta_0,R_1(x^*_{i1};h_1)+\frac{\delta_1}{h_1},R_2(x^*_{i1};h_1)+\frac{\delta_2}{h_1^2}\right)\nonumber\\
   &=\frac{1}{nh_1}K\left(\frac{x^*_{j1}-x^*_{i1}}{h_1}\right)Q_2(x^*_{i1};h_1)\left\{1+O_{UP}\left(\frac{1}{qh_1^2}\right)\right\}\quad\quad\quad\quad\quad\quad\quad\quad\quad\text{\{by Lemma S\ref{f_t}\}}\nonumber\\
    &\indent-\frac{1}{nh_1}K\left(\frac{x^*_{j1}-x^*_{i1}}{h_1}\right)\frac{x^*_{j1}-x^*_{i1}}{h_1}Q_1(x^*_{i1};h_1)\left\{1+O_{UP}\left(\frac{1}{qh_1^2}\right)\right\}\nonumber\\
    &=\left\{\frac{1}{nh_1}K\left(\frac{x^*_{j1}-x^*_{i1}}{h_1}\right)Q_2(x^*_{i1};h_1)-\frac{1}{nh_1}K\left(\frac{x^*_{j1}-x^*_{i1}}{h_1}\right)\frac{x^*_{j1}-x^*_{i1}}{h_1}Q_1(x^*_{i1};h_1)\right\}\nonumber\\
&\indent\times\left\{1+O_{UP}\left(\frac{1}{qh_1^2}\right)\right\}.
\end{align}
\noindent
Similarly we obtain
\begin{align}\label{s2ij}
    [\mathcal{S}_2]_{ij} &= \left\{\frac{1}{nh_2}K\left(\frac{x^*_{j2}-x^*_{i2}}{h_2}\right)Q_2(x^*_{i2};h_2)-\frac{1}{nh_2}K\left(\frac{x^*_{j2}-x^*_{i2}}{h_2}\right)\frac{x^*_{j2}-x^*_{i2}}{h_2}Q_1(x^*_{i2};h_2)\right\}\nonumber\\
    &\indent\times\left\{1+O_{UP}\left(\frac{1}{qh_2^2}\right)\right\}.
\end{align}
Since 
$\mathcal{S}_t^*=(\mathcal{I}-\mathbf{1}\mathbf{1}^T/n)\mathcal{S}_t$ for $t=1,2$, we have
\begin{align*}
[\mathcal{S}_t^*]_{ij}&=[\mathcal{S}_t]_{ij}-\frac{1}{n}\sum_{l=1}^{n}[\mathcal{S}_t]_{lj},\quad t=1, 2.
\end{align*}
Then 
\begin{align}
[\mathcal{S}_1^*\mathcal{S}_2^*]_{ij}&=\sum_{m=1}^{n}\left([\mathcal{S}_1]_{im}[\mathcal{S}_2]_{mj}+\frac{1}{n^2}\sum_{l=1}^{n}[\mathcal{S}_1]_{lm}\sum_{l'=1}^{n}[\mathcal{S}_2]_{l'j}-\frac{1}{n}[\mathcal{S}_2]_{mj}\sum_{l=1}^{n}[\mathcal{S}_1]_{lm}-\frac{1}{n}[\mathcal{S}_1]_{im}\sum_{l'=1}^{n}[\mathcal{S}_2]_{l'j}\right)\nonumber\\
    &=\left\{[\mathcal{S}_1\mathcal{S}_2]_{ij} -\frac{1}{n}\sum_{m=1}^{n}[\mathcal{S}_1]_{im}\sum_{l'=1}^{n}[\mathcal{S}_2]_{l'j}\right\}(\text{denoted by } L_1)
    \nonumber
    \\
    &\indent+\left[\frac{1}{n}\left\{\frac{1}{n}\sum_{m=1}^{n}\sum_{l=1}^{n}[\mathcal{S}_1]_{lm}\sum_{l'=1}^{n}[\mathcal{S}_2]_{l'j}-\sum_{m=1}^{n}\left([\mathcal{S}_2]_{mj}\sum_{l=1}^{n}[\mathcal{S}_1]_{lm}\right)\right\} \right](\text{denoted by }L_2)
    \label{trInt}.
\end{align}
We next prove that both $L_1=o_{UP}(1/n)$ and $L_2=o_{UP}(1/n)$ for large enough $n$ so that $h_1,h_2\le1/2$.


\medskip
\noindent
\textbf{Proof that $L_1=o_{UP}(1/n)$.}
By \eqref{s1ij} and \eqref{s2ij}, for the first term in $L_1$, we have 
\begin{align}\label{eq:s12}
[\mathcal{S}_1\mathcal{S}_2]_{ij}&=\left\{\frac{1}{n^2h_1h_2}\sum_{l=1}^{n}K\left(\frac{x^*_{l1}-x^*_{i1}}{h_1}\right)Q_2(x^*_{i1};h_1)K\left(\frac{x^*_{j2}-x^*_{l2}}{h_2}\right)Q_2(x^*_{l2};h_2)\right.\nonumber\\
    &+\frac{1}{n^2h_1h_2}\sum_{l=1}^{n}K\left(\frac{x^*_{l1}-x^*_{i1}}{h_1}\right)\frac{x^*_{l1}-x^*_{i1}}{h_1}Q_1(x^*_{i1};h_1)K\left(\frac{x^*_{j2}-x^*_{l2}}{h_2}\right)\frac{x^*_{j2}-x^*_{l2}}{h_2}Q_1(x^*_{l2};h_2)\nonumber\\
    &-\frac{1}{n^2h_1h_2}\sum_{l=1}^{n}K\left(\frac{x^*_{l1}-x^*_{i1}}{h_1}\right)Q_2(x^*_{i1};h_1)K\left(\frac{x^*_{j2}-x^*_{l2}}{h_2}\right)\frac{x^*_{j2}-x^*_{l2}}{h_2}Q_1(x^*_{l2};h_2)\nonumber\\
    &\left.-\frac{1}{n^2h_1h_2}\sum_{l=1}^{n}K\left(\frac{x^*_{l1}-x^*_{i1}}{h_1}\right)\frac{x^*_{l1}-x^*_{i1}}{h_1}Q_1(x^*_{i1};h_1)K\left(\frac{x^*_{j2}-x^*_{l2}}{h_2}\right)Q_2(x^*_{l2};h_2)\right\}\nonumber\\
    &\times\left\{1+O_{UP}\left(\frac{1}{qh_1^2}+\frac{1}{qh_2^2}\right)\right\}.
\end{align}
By Lemma S\ref{f_t},  $Q_t(x;h)$ is a $(C_t/h)$-Lipschitz function for some constant $C_t, \ t=1,2$. By Lemma S\ref{Q} and Assumption \ref{a5}, when $n$ is large, $\sup_{h_1\in(0,1/2]}\sup_{x\in[0,1]}|Q_t(x;h)|$ is bounded above by a constant. Therefore, we can apply Lemma S\ref{2driemannintegral} to \eqref{eq:s12} to obtain
\begin{align}\label{s12int}
    [\mathcal{S}_1\mathcal{S}_2]_{ij} &=\frac{1}{nh_1h_2}\left\{\int_{0}^{1}K\left(\frac{x-x^*_{i1}}{h_1}\right)Q_2(x^{*}_{i1};h_1)dx\int_{0}^{1}K\left(\frac{x^*_{j2}-x'}{h_2}\right)Q_2(x';h_2)dx'\right.\nonumber\\
    &+\int_{0}^{1}K\left(\frac{x-x^*_{i1}}{h_1}\right)\frac{x-x^*_{i1}}{h_1}Q_1(x^{*}_{i1};h_1)dx\int_{0}^{1}K\left(\frac{x^*_{j2}-x'}{h_2}\right)\frac{x^*_{j2}-x'}{h_2}Q_1(x';h_2)dx'\nonumber\\
    &-\int_{0}^{1}K\left(\frac{x-x^*_{i1}}{h_1}\right)Q_2(x^{*}_{i1};h_1)dx\int_{0}^{1}K\left(\frac{x^*_{j2}-x'}{h_2}\right)\frac{x^*_{j2}-x'}{h_2}Q_1(x';h_2)dx'\nonumber\\
    &-\int_{0}^{1}K\left(\frac{x-x^*_{i1}}{h_1}\right)\frac{x-x^*_{i1}}{h_1}Q_1(x^{*}_{i1};h_1)dx\int_{0}^{1}K\left(\frac{x^*_{j2}-x'}{h_2}\right)Q_2(x';h_2)dx'\nonumber\\
    &\left.+O_{UP}\left(\frac{1}{qh_1}+\frac{1}{qh_2}\right)\right\}\left\{1+O_{UP}\left(\frac{1}{qh_1^2}+\frac{1}{qh_2^2}\right)\right\}\nonumber\\
    &=\frac{1}{n}\left\{\int_{\mathcal{D}_{x^*_{i1},h_1}}K\left(u\right)Q_2(x^{*}_{i1};h_1)du\int_{\mathcal{D}_{x^*_{j2},h_2}}K\left(u'\right)Q_2(x^*_{j2}+u'h_2;h_2)du'\right.\nonumber\\
    &-\int_{\mathcal{D}_{x^*_{i1},h_1}}K\left(u\right)uQ_1(x^{*}_{i1};h_1)du\int_{\mathcal{D}_{x^*_{j2},h_2}}K\left(u'\right)uQ_1(x^*_{j2}+u'h_2;h_2)du'\nonumber\\
    &+\int_{\mathcal{D}_{x^*_{i1},h_1}}K\left(u\right)Q_2(x^{*}_{i1};h_1)du\int_{\mathcal{D}_{x^*_{j2},h_2}}K\left(u'\right)u'Q_1(x^*_{j2}+u'h_2;h_2)du'\nonumber\\
    &-\int_{\mathcal{D}_{x^*_{i1},h_1}}K\left(u\right)uQ_1(x^{*}_{i1};h_1)du\int_{\mathcal{D}_{x^*_{j2},h_2}}K\left(u'\right)Q_2(x^*_{j2}+u'h_2;h_2)du'\nonumber\\
    &\left.+O_{UP}\left(\frac{1}{qh_1^2h_2}+\frac{1}{qh_1h_2^2}\right)\right\}\left\{1+O_{UP}\left(\frac{1}{qh_1^2}+\frac{1}{qh_2^2}\right)\right\},
\end{align}
where the last equality is obtained by letting $u=(x-x^*_{i1})/h_1$ and $u'=(x'-x^*_{j2})/h_2$.

For the second term in $L_1$, we apply Lemma S\ref{1dRieSum} to obtain
\begin{align}\label{fi}
    \sum_{m=1}^{n}[\mathcal{S}_1]_{im}&=\frac{1}{h_1}\left\{\int_{0}^{1}K\left(\frac{x-x^{*}_{i1}}{h_1}\right)Q_2(x^{*}_{i1};h_1)dx\right.\nonumber\\
    &\left.-\int_{0}^{1}K\left(\frac{x-x^{*}_{i1}}{h_1}\right)\frac{x-x^{*}_{i1}}{h_1}Q_1(x^{*}_{i1};h_1)dx+O_{UP}\left(\frac{1}{qh_1}\right)\right\}\left\{1+O_{UP}\left(\frac{1}{qh_1^2}\right)\right\}\nonumber\\
    &=\left\{\int_{\mathcal{D}_{x^*_{i1},h_1}}K\left(u\right)Q_2(x^{*}_{i1};h_1)du\right.\nonumber\\
    &\left.-\int_{\mathcal{D}_{x^*_{i1},h_1}}K\left(u\right)uQ_1(x^{*}_{i1};h_1)du+O_{UP}\left(\frac{1}{qh_1^2}\right)\right\}\left\{1+O_{UP}\left(\frac{1}{qh_1^2}\right)\right\},
\end{align}
and
\begin{align}\label{fj}
    \sum_{l'=1}^{n}[\mathcal{S}_2]_{l'j}&=\frac{1}{h_2}\left\{\int_{0}^{1}K\left(\frac{x^{*}_{j2}-x'}{h_2}\right)Q_2(x';h_2)dx'\right.\nonumber\\
    &\left.-\int_{0}^{1}K\left(\frac{x^{*}_{j2}-x'}{h_2}\right)\frac{x^{*}_{j2}-x'}{h_2}Q_1(x';h_2)dx'+O_{UP}\left(\frac{1}{qh_2}\right)\right\}\left\{1+O_{UP}\left(\frac{1}{qh_2^2}\right)\right\}\nonumber\\
    &=\left\{\int_{\mathcal{D}_{x^*_{j2},h_2}}K\left(u'\right)Q_2(x^*_{j2}+u'h_2;h_2)du'\right.\nonumber\\
    &\left.+\int_{\mathcal{D}_{x^*_{j2},h_2}}K\left(u'\right)u'Q_1(x^*_{j2}+u'h_2;h_2)du'+O_{UP}\left(\frac{1}{qh_2^2}\right)\right\}\left\{1+O_{UP}\left(\frac{1}{qh_2^2}\right)\right\}.
\end{align}
Combing \eqref{s12int},
\eqref{fi} and \eqref{fj}, we have
\begin{align}\label{s20} 
L_1=\frac{1}{n}O_{UP}\left(\frac{1}{qh_1^2}+\frac{1}{qh_2^2}\right) = o_{UP}\left(\frac{1}{n} \right),
\end{align}
since $1/(qh_1^2)\rightarrow 0$ and $1/(qh_2^2)\rightarrow 0$ by Assumptions \ref{a1} and \ref{a5}.


\medskip
\noindent
\textbf{Proof that $L_2=o_{UP}(1/n)$.}
We first evaluate the first term in $L_2$. By \eqref{fj}, we have 
\begin{align*}
    \sum_{l=1}^{n}[\mathcal{S}_1]_{lm}&=\frac{1}{h_1}\left\{\mathbb{L}(x^{*}_{m1};h_1)+O_{UP}\left(\frac{1}{qh_1}\right)\right\}\left\{1+O_{UP}\left(\frac{1}{qh_1^2}\right)\right\},
\end{align*}
where 
\begin{align*}
 \mathbb{L}(x^{*}_{m1};h_1)&:=\int_{0}^{1}K\left(\frac{x^{*}_{m1}-\nu}{h_1}\right)Q_2(\nu;h_1)d\nu\nonumber
    -\int_{0}^{1}K\left(\frac{x^{*}_{m1}-\nu}{h_1}\right)\frac{x^{*}_{m1}-\nu}{h_1}Q_1(\nu;h_1)d\nu.   
\end{align*}
Then $$\frac{1}{n}\sum_{m=1}^{n}\sum_{l=1}^{n}[\mathcal{S}_1]_{lm}=\frac{1}{h_1}\left\{\frac{1}{n}\sum_{m=1}^{n}\mathbb{L}(x^{*}_{m1};h_1)+O_{UP}\left(\frac{1}{qh_1}\right)\right\}\left\{1+O_{UP}\left(\frac{1}{qh_1^2}\right)\right\}.$$
Now we work on $(1/n)\sum_{m=1}^{n}\mathbb{L}(x^{*}_{m1};h_1)$.  By Lemma S\ref{Q}, $\sup_{h\in(0,1/2]}\sup_{x\in[0,1]}|Q_t(x;h)|$ is bounded above by some constant, say $\xi_t$, for $t=1,2$, which implies that \\
$\sup_{h\in(0,1/2]}\sup_{x\in[0,1]}\left|\mathbb{L}(x;h)\right| \leq M_0(\xi_1+\xi_2)$. In addition, $\mathbb{L}(x;h)/h$ is $\{(\kappa_1+\kappa_2)/h\}$-Lipschitz by Lemma S\ref{intlip}. Thus, by another Riemann integral approximation with respect to variable $x^{*}_{m1}$, we have 
\begin{align*}
  \frac{1}{n}\sum_{m=1}^{n}\mathbb{L}(x^{*}_{m1};h_1)=&\int_{0}^{1}\int_{0}^{1}\left\{K\left(\frac{x-\nu}{h_1}\right)Q_2(\nu;h_1)-K\left(\frac{x-\nu}{h_1}\right)\frac{x-\nu}{h_1}Q_1(\nu;h_1)\right\}dxd\nu\\
  &+O_{UP}\left(\frac{1}{qh_1}\right),  
\end{align*}
which implies that 
\begin{align}\label{s1dbint}
    \frac{1}{n}\sum_{m=1}^{n}\sum_{l=1}^{n}[\mathcal{S}_1]_{lm}&=\frac{1}{h_1}\left[\int_{0}^{1}\int_{0}^{1}\left\{K\left(\frac{x-\nu}{h_1}\right)Q_2(\nu;h_1)-K\left(\frac{x-\nu}{h_1}\right)\frac{x-\nu}{h_1}Q_1(\nu;h_1)\right\}dxd\nu\right.\nonumber\\
    &\indent\left.+O_{UP}\left(\frac{1}{qh_1}\right)\right]
    \left\{1+O_{UP}\left(\frac{1}{qh_1^2}\right)\right\}.
\end{align}
With \eqref{fi} and \eqref{s1dbint}, we have 
\begin{align}\label{sss}
    &\indent\frac{1}{n}\sum_{m=1}^{n}\sum_{l=1}^{n}[\mathcal{S}_1]_{lm}\sum_{l'=1}^{n}[\mathcal{S}_2]_{l'j}=\frac{1}{h_1h_2}\left\{1+O_{UP}\left(\frac{1}{qh_1^2}\right)+O_{UP}\left(\frac{1}{qh_2^2}\right)\right\}\times\nonumber\\
    &\quad\left[\int_{0}^{1}\int_{0}^{1}\left\{K\left(\frac{x-\nu}{h_1}\right)Q_2(\nu;h_1)-K\left(\frac{x-\nu}{h_1}\right)\frac{x-\nu}{h_1}Q_1(\nu;h_1)\right\}dxd\nu
    +O_{UP}\left(\frac{1}{qh_1}\right)\right]\times\nonumber\\
    &\quad\left\{\int_{0}^{1}K\left(\frac{x^{*}_{j2}-x'}{h_2}\right)Q_2(x';h_2)dx'-\int_{0}^{1}K\left(\frac{x^{*}_{j2}-x'}{h_2}\right)\frac{x^{*}_{j2}-x'}{h_2}Q_1(x';h_2)dx+O_{UP}\left(\frac{1}{qh_2}\right)\right\}\nonumber\\
    &=\left[\int_{0}^{1}\int_{\mathcal{D}_{v,h_1}}\left\{K\left(u\right)Q_2(\nu;h_1)-K\left(u\right)uQ_1(\nu;h_1)\right\}dud\nu
    +O_{UP}\left(\frac{1}{qh_1^2}\right)\right]\times\nonumber\\
    &\quad\left[\int_{\mathcal{D}_{x^*_{j2},h_2}}\left\{K\left(u'\right)Q_2(x^*_{j2}+u'h_2;h_2)+K\left(u'\right)u'Q_1(x_{j2}^{*}+u'h_2;h_2)\right\}du'+O_{UP}\left(\frac{1}{qh_2^2}\right)\right]\nonumber\\
    &\indent\times\left\{1+O_{UP}\left(\frac{1}{qh_1^2}\right)+O_{UP}\left(\frac{1}{qh_2^2}\right)\right\},
\end{align}
where the last equality is obtained by letting $u=(x-\nu)/h_1$ and $u'=\left(x'-x^{*}_{j2}\right)/h_2$.


For the second term in $L_2$, by \eqref{fj}, 
\begin{align}
    &\sum_{m=1}^{n}\left([\mathcal{S}_2]_{mj}\sum_{l=1}^{n}[\mathcal{S}_1]_{lm}\right)=\frac{1}{nh_1h_2}\left\{1+O_{UP}\left(\frac{1}{qh_1^2}\right)+O_{UP}\left(\frac{1}{qh_2^2}\right)\right\}\times\nonumber\\
    &\sum_{m=1}^{n}\left[\left\{\int_{0}^{1}K\left(\frac{\nu-x^{*}_{m1}}{h_1}\right)Q_2(\nu;h_1)+K\left(\frac{\nu-x^{*}_{m1}}{h_1}\right)\frac{\nu-x^{*}_{m1}}{h_1}Q_1(\nu;h_1)d\nu+O_{UP}\left(\frac{1}{qh_1}\right)\right\}\right.\nonumber\\
&\left.\left\{K\left(\frac{x^*_{j2}-x^*_{m2}}{h_2}\right)Q_2(x^*_{m2};h_2)-K\left(\frac{x^*_{j2}-x^*_{m2}}{h_2}\right)\frac{x^*_{j2}-x^*_{m2}}{h_2}Q_1(x^*_{m2};h_2)\right\}\right],\label{trInt2}
\end{align}
which can be considered as a two-dimensional Riemann sum over $\left(x^{*}_{m1},x^{*}_{m2}\right)$ for $m=1,\dots,n$. Similar to the proof of Lemma S\ref{2driemannintegral}, we have
\begin{align}\label{trInt3}
    &\sum_{m=1}^{n}\left([\mathcal{S}_2]_{mj}\sum_{l=1}^{n}[\mathcal{S}_1]_{lm}\right)=\frac{1}{h_1h_2}\left\{1+O_{UP}\left(\frac{1}{qh_1^2}\right)+O_{UP}\left(\frac{1}{qh_2^2}\right)\right\}\times\nonumber\\
    &\indent\left\{\int_{0}^{1}\int_{0}^{1}K\left(\frac{\nu-x}{h_1}\right)Q_2(\nu;h_1)+K\left(\frac{\nu-x}{h_1}\right)\frac{\nu-x}{h_1}Q_1(\nu;h_1)dx d\nu +O_{UP}\left(\frac{1}{qh_1^2}\right)\right\}\nonumber\\
    &\indent\left\{\int_{0}^{1}K\left(\frac{x^{*}_{j2}-x'}{h_2}\right)Q_2(x';h_2)-K\left(\frac{x^{*}_{j2}-x}{h_2}\right)\frac{x^{*}_{j2}-x'}{h_2}Q_1(x';h_2)dx'+O_{UP}\left(\frac{1}{qh_2^2}\right) \right\}\nonumber\\&=\left[\int_{0}^{1}\int_{\mathcal{D}_{v,h_1}}\left\{K\left(u\right)Q_2(\nu;h_1)-K\left(u\right)uQ_1(\nu;h_1)\right\}dud\nu
    +O_{UP}\left(\frac{1}{qh_1^2}\right)\right]\times\nonumber\\
    &\quad\left[\int_{\mathcal{D}_{x^*_{j2},h_2}}\left\{K\left(u'\right)Q_2(x^*_{j2}+u'h_2;h_2)+K\left(u'\right)u'Q_1(x_{j2}^{*}+u'h_2;h_2)\right\}du'+O_{UP}\left(\frac{1}{qh_2^2}\right)\right]\nonumber\\
    &\indent\times\left\{1+O_{UP}\left(\frac{1}{qh_1^2}\right)+O_{UP}\left(\frac{1}{qh_2^2}\right)\right\},
\end{align}
where in the last step we let $u=(x-\nu)/h_1$ and $u'=(x'-x^{*}_{j2})/h_2$. 
Subtracting \eqref{trInt3} from \eqref{sss}, we obtain
$$
L_2=\frac{1}{n}O_{UP}\left(\frac{1}{qh_1^2}+\frac{1}{qh_2^2}\right)=o_{UP}\left(\frac{1}{n} \right),$$
since $1/(qh_1^2)\rightarrow0$ and $1/(qh_2^2)\rightarrow0$ by Assumptions \ref{a1} and \ref{a5}.

Since $[\mathcal{S}_1^*\mathcal{S}_2^*]_{ij}=L_1+L_2$ by \eqref{trInt}, we have 
$
[\mathcal{S}_1^*\mathcal{S}_2^*]_{ij}=o_{UP}\left(n^{-1} \right).
$
Therefore, under $E_N$, of which probability approaches to one as $N\rightarrow \infty$, we have
$$
\|\mathcal{S}_1^*\mathcal{S}_2^*\|_{\infty}=o_{UP}\left(1\right).$$
This completes the proof. 
\end{proof}

\section{Additional Simulation Results}
\begin{figure}[H]
  \centering \includegraphics[width=1\linewidth]{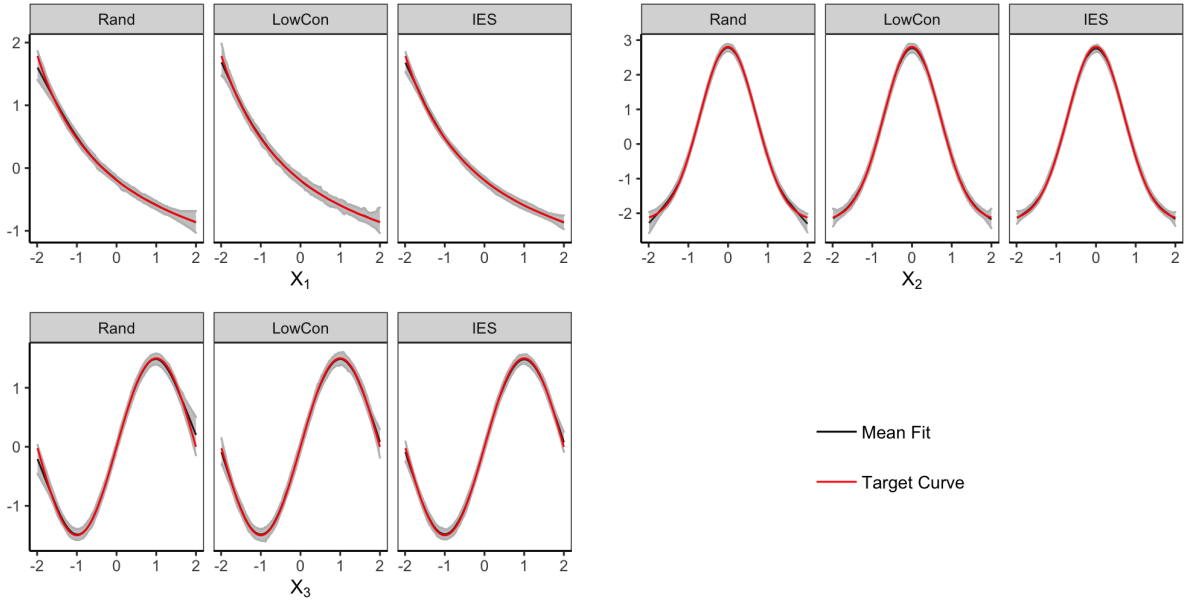}
\caption{Component function estimates trained on subsamples obtained by different methods for Case 1 (truncated multivariate normal predictors) in Section \ref{sec:simu}.}
\label{fig:normcompare}
\end{figure}

\begin{figure}[H]
  \centering \includegraphics[width=1\linewidth]{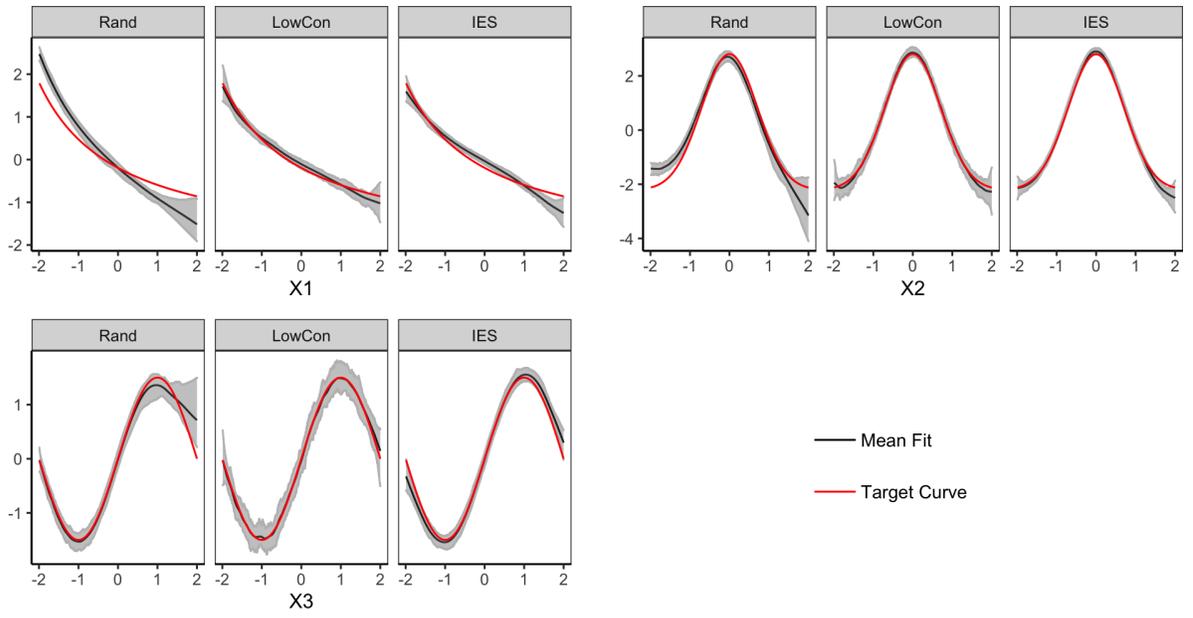}
\caption{Component function estimates trained on subsamples obtained by different methods for exponential predictors under model misspecification.}
\label{fig:miscompare}
\end{figure}

Figure \ref{fig:normcompare} plots the component function estimates trained on subsamples obtained by different methods for Case 1 ( truncated multivariate normal predictors) in Section \ref{sec:simu}. Figure \ref{fig:miscompare} plots the component function estimates trained on subsamples obtained by different methods for truncated multivariate exponential predictors under model misspecification. Details are discussed in Section \ref{sec:simu}.

\noindent

\newpage
\bibliographystyle{chicago}
\bibliography{Bography}
\makeatletter\@input{withSupp.tex}\makeatother